\documentclass[10pt]{article}

\usepackage[T1]{fontenc}
\usepackage[utf8]{inputenc}
\usepackage{arxiv}

\usepackage{amsmath}
\usepackage{amssymb}
\usepackage{graphicx}
\usepackage{xcolor}
\usepackage{booktabs}
\usepackage{placeins}
\usepackage{subcaption}
\usepackage[numbers]{natbib}
\usepackage{microtype}
\usepackage{etoolbox}
\usepackage{hyperref}
\hypersetup{
  hidelinks,
  pdftitle={GeoCANO: A Geometry-Conditioned Adaptive Neural Operator for Physical-Field Prediction on Varying Domains},
  pdfauthor={Biao Chen, Xinrui Zhou, Xianlin Jin, Jing Wang, Jifa Zhang, Chen Zhai, Wei Liu, Miao Zhang},
  pdfkeywords={Neural operator, Physics field prediction, Mixture of experts, Transonic aerodynamics}
}
\renewcommand{\shorttitle}{GeoCANO}
\date{}
\makeatletter
\patchcmd{\@maketitle}
  {\begin{tabular}[t]{c}\bf\rule{\z@}{24\p@}\@author\end{tabular}}
  {\ifx\@author\@empty\else
     \begin{tabular}[t]{c}\bf\rule{\z@}{24\p@}\@author\end{tabular}\fi}
  {}{\PackageError{geocano-arxiv}{Author-layout patch failed}{Check arxiv.sty.}}
\patchcmd{\@maketitle}
  {\vskip 0.4in \@minus 0.1in \center{\@date} \vskip 0.2in}
  {\ifx\@date\@empty\vskip 0.15in\else
     \vskip 0.4in \@minus 0.1in \center{\@date} \vskip 0.2in\fi}
  {}{\PackageError{geocano-arxiv}{Date-layout patch failed}{Check arxiv.sty.}}
\makeatother

\begin{document}

\title{GeoCANO: A Geometry-Conditioned Adaptive Neural Operator for Physical-Field Prediction on Varying Domains}

\author{%
  \textbf{Biao Chen\textsuperscript{1} \quad
  Xinrui Zhou\textsuperscript{1} \quad
  Xianlin Jin\textsuperscript{2} \quad
  Jing Wang\textsuperscript{1}}\\[4pt]
  \textbf{Jifa Zhang\textsuperscript{3} \quad
  Chen Zhai\textsuperscript{4} \quad
  Wei Liu\textsuperscript{4} \quad
  Miao Zhang\textsuperscript{4}}\\[6pt]
  {\normalfont\small\textsuperscript{1}Shanghai Jiao Tong University}\\
  {\normalfont\small\textsuperscript{2}Huawei Technologies Co Ltd}\\
  {\normalfont\small\textsuperscript{3}Zhejiang University}\\
  {\normalfont\small\textsuperscript{4}Shanghai Aircraft Design and Research Institute}%
}
\maketitle

\begin{abstract}
Industrial design requires repeated evaluation of candidate configurations over large design spaces. Integral performance measures alone do not resolve critical local phenomena, whereas high-fidelity simulation of every configuration remains too expensive for rapid design iteration. Existing physical-field surrogates commonly rely on prescribed output discretizations, coupling geometry representation and coordinate evaluation through mesh topology and thereby limiting reuse across meshes of different resolutions. We propose the Geometry-Conditioned Adaptive Neural Operator (GeoCANO), which casts physical-field prediction over varying three-dimensional geometries as a mapping from a sampled complete geometry to a coordinate-dependent field function. An attention-based encoder extracts a geometry state from unordered surface coordinates. To enhance multiscale expressiveness, we introduce a multiresolution axis-factorized hash encoding together with a query-wise sparse mixture-of-experts decoder. The model constructs the geometry-conditioned field representation before introducing the output coordinates, enabling a single geometry encoding to be evaluated directly in different regions without rebuilding the computational graph associated with the output topology. We evaluated GeoCANO on transonic aerodynamic-field prediction over complete aircraft configurations using a Reynolds-averaged Navier--Stokes dataset comprising 10,484 wing--body--nacelle geometries. For unseen configurations within the sampled design family, GeoCANO outperformed the baseline models in complete-surface prediction, high-gradient pressure reconstruction, and pressure-drag recovery. Relative to the strongest baseline, it reduced the mean absolute error in the pressure coefficient by 77.9\% and retained a clear advantage over a finite range of wall-normal distances. Ablation studies showed that the geometry-modulated coordinate representation achieved comparable accuracy with substantially fewer parameters than conventional three-dimensional hash encoding. Routing analysis further indicated that the sparse experts developed spatially structured, query-dependent specialization. These results demonstrate that GeoCANO provides a reusable continuous-coordinate representation for rapid physical-field evaluation over varying geometries.

\end{abstract}


\section{Introduction}

Engineering design problems in automotive \cite{vatani2025tripoptimizer}, marine \cite{serani2024marine}, and aircraft configuration design \cite{leclainche2023review,immordino2024steady} commonly require the repeated evaluation of candidate designs over large geometric design spaces. Because integral performance measures alone cannot reveal the spatial origins of performance variations or critical local phenomena, engineering assessment also requires spatially resolved physical fields. High-fidelity numerical simulations provide such information, but repeated preprocessing and converged solutions for numerous candidate geometries incur computational costs that are often incompatible with rapid design iteration. Under prescribed operating and boundary conditions, surrogate models, particularly neural operators, learn mappings from complete geometries to their associated physical fields \cite{li2023gkn,azizzadenesheli2024review}, providing an effective means of reducing repeated evaluation costs and enabling rapid field prediction for complex physical systems \cite{pestourie2023physicsenhanced,howard2023multifidelity,kontolati2024latent}.

Learning such geometry-to-field mappings first requires different computational domains to be represented within a common framework. Structured-grid convolutional models \cite{zuo2024fast} and reduced-order methods \cite{oleary2024dino} generally rely on a shared spatial parameterization or a stable correspondence between discrete degrees of freedom, assumptions that may be invalidated by geometric variation. Graph models \cite{pfaff2021meshgraphnets,pichi2024graph,immordino2025gcn}, point-cloud models \cite{he2024geomdeeponet,wang2025pointcloud}, and attention architectures for unstructured discretizations \cite{wu2024transolver} operate directly on geometric samples and thereby reduce the dependence on regular grids and explicit parameterizations. For complex computational domains, existing neural operators extend operator learning through latent regular grids, learned geometric deformations, or point-cloud integration \cite{li2023gino,li2023geofno,zeng2025pcno,liu2026ginot}. These approaches relax the requirement for a common regular representation and enable physical-field prediction across geometries. However, the predicted fields are still commonly generated on sample-specific output discretizations, such that the output point sets and their topological relations remain part of the predictive representation. Consequently, the ability to accommodate geometric variation does not necessarily imply independence from the output discretization.

This dependence on the output discretization conflicts with the requirements of industrial design. The same configuration may be assessed on meshes with different resolutions or local refinements, and post-processing may require field values at coordinates absent from the original mesh. Most high-accuracy models for unstructured domains represent both the input geometry and the output field as finite point sets. Their output nodes therefore play a dual role: they carry the predicted field values while also participating in information propagation and local feature construction, as in multimesh graph networks \cite{massegur2024multimesh} and point-cloud models \cite{wang2025pointcloud}. Changing the mesh consequently alters not only the locations at which the field is sampled but also the internal information pathways of the model. Such models reconstruct a field on a prescribed discrete structure rather than define a field determined by the complete geometry and independently evaluable at arbitrary coordinates. From a physical standpoint, different meshes are merely alternative discretizations of the same state; they should neither alter predictions at shared coordinates nor require the complete geometry to be encoded again. Variable-geometry field modeling must therefore remove the coupling between geometry representation and output discretization, so that a single geometry encoding supports consistent evaluation on multiple meshes and at arbitrary coordinates.

Decoding the physical field as a coordinate-dependent function provides a direct route to this separation. The complete geometry is first mapped to a conditional representation that is independent of the output mesh, after which field values are recovered using spatial coordinates as queries. DeepONet \cite{lu2021deeponet} established the basic form of coordinate-conditioned operator learning, while continuous neural fields \cite{catalani2024neuralfields} have enabled coordinate-wise reconstruction on complex surfaces. For varying geometries, reference-domain mappings \cite{yin2024dimon}, continuous geometry-aware reduced-order representations \cite{brivio2025cga}, and implicit shape encodings \cite{zhang2025shape} connect geometric variation to continuous fields from complementary perspectives. Geometry-conditioned operator formulations \cite{zhong2025pigano,peyvan2026fusion} further demonstrate the feasibility of pointwise prediction on complex three-dimensional configurations. These advances show that geometry encoding and coordinate evaluation can be separated structurally. Their accuracy, however, remains governed by the capacity of the reusable geometry representation to retain the complete physical state. When the full geometry is compressed into a compact global code and all locations are recovered by a shared coordinate decoder, localized complex structures can be limited by the resulting representational bottleneck.

This difficulty arises from the scale mismatch between global geometric control and local field recovery. Geometric changes alter boundary locations and component relationships, thereby reorganizing the physical state globally, whereas coordinate-based functional representations commonly employ a globally shared decoding mechanism. The spatial complexity of an engineering field is nevertheless highly nonuniform: the field varies smoothly over much of the domain, while steep-gradient structures occupy only limited regions. Local neural operators \cite{li2024local,ye2024locality} show that finite receptive fields can capture local regularities on varying domains, and low-rank representations \cite{venturi2022svd,kolda2009tensor,chen2022tensorf} reduce the cost of high-dimensional field representation and repeated evaluation. A fixed low-rank space, however, cannot readily reallocate its representational capacity as geometry changes. Coordinate encodings provide a complementary mechanism for resolving fine-scale structure. Fourier features \cite{tancik2020fourier} enhance the high-frequency expressiveness of coordinate networks, while multiresolution hash encodings \cite{muller2022hash,huang2024hash} organize multiple spatial scales at low computational cost. For transonic-field prediction, PHD-NO \cite{chen2026phdno} further shows that assigning distinct representations to the background field and shock structures improves reconstruction accuracy. Nevertheless, the regions requiring high representational capacity move with the geometry, and fixed coordinate features cannot determine a priori where model capacity should be concentrated. The spectral bias of neural networks \cite{rahaman2019spectral}, which remains evident in multiscale operator learning \cite{liu2024spectral}, further accentuates this limitation.

Based on this analysis, we propose the \emph{Geometry-Conditioned Adaptive Neural Operator} (GeoCANO), which formulates physical-field prediction over varying three-dimensional geometries as a continuous approximation conditioned on the complete geometry. The model first extracts a global state from unordered geometric samples. Its latent-attention encoder follows the set-based formulation used in point-cloud representation learning \cite{pang2022pointmae}, making the geometry encoding invariant to node ordering. The global state then modulates a low-rank multiresolution representation shared across configurations, establishing a fixed field approximation space for the current geometry before any output coordinates are introduced. This space is constructed from an axis-factorized canonical polyadic representation, which organizes multiscale field information with modest storage and evaluation costs. Given a query coordinate, a local evaluator recovers the corresponding field value solely within the conditioned approximation space. A sparse mixture-of-experts mechanism \cite{shazeer2017outrageously} selects an appropriate local approximation according to the positional features without altering the global geometry state. A single geometry encoding can consequently serve different resolutions, local regions, and query batches, while predictions at shared coordinates remain independent of the other queried points. Global geometry modulation and query-dependent local adaptation thus form a unified conditional approximation that combines query decoupling with the resolution of localized high-gradient structures.

We consider transonic surface-pressure prediction for complete aircraft configurations as a representative application. In this problem, small geometric changes can displace shocks and modify wing--body--nacelle interference, giving rise to both global geometric effects and localized steep gradients \cite{hines2023graph}. Previous studies have shown that data-driven models can reconstruct aerodynamic fields on complex aircraft geometries. This application therefore provides a stringent setting in which to assess the query consistency and field-reconstruction accuracy of GeoCANO under strong geometric sensitivity. The principal contributions of this work are as follows:

\begin{enumerate}
    \item We introduce a framework that separates geometry encoding from coordinate-wise evaluation for physical fields on varying three-dimensional domains. A field representation conditioned on the complete geometry is established before querying, allowing the same geometry encoding to be reused across output discretizations while preserving consistent predictions at shared coordinates.
    \item We construct a low-rank multiresolution approximation space modulated by the global geometry and combine it with query-dependent local expert evaluation. This formulation resolves spatially nonuniform physical structures while retaining a reusable geometry representation.
    \item We evaluate GeoCANO on a transonic Reynolds-averaged Navier--Stokes dataset comprising 10,484 wing--body--nacelle configurations. The experiments assess pressure reconstruction on different surface discretizations, pressure-drag recovery, and direct evaluation on successive off-wall layers.
\end{enumerate}

The remainder of this paper is organized as follows. Section~\ref{sec:method} formulates the learning problem and presents the GeoCANO architecture and training procedure. Section~\ref{sec:dataset} describes the geometry family, CFD dataset construction, and data partition. Section~\ref{sec:experiments} reports the predictive results and engineering assessments. Section~\ref{sec:conclusion} concludes the paper.

\section{GeoCANO: A Geometry-Conditioned Adaptive Neural Operator for Continuous-Coordinate Surface-Field Prediction}
\label{sec:method}

\subsection{Geometry-conditioned operator formulation}
\label{sec:operator-formulation}
Let $\Gamma\subset\mathbb{R}^{d}$ denote a geometry-dependent domain and
$\mathbf{y}:\Gamma\rightarrow\mathbb{R}^{C}$ a field defined on that domain.
A learning sample provides a finite geometric observation
$X=\{\mathbf{x}_{i}\}_{i=1}^{N_e}\subset\Gamma$ and a finite query set
$Q=\{\mathbf{q}_{j}\}_{j=1}^{N_q}\subset\Gamma$. The corresponding discrete
field is
$\mathbf{Y}(Q)=[\mathbf{y}(\mathbf{q}_{j})]_{j=1}^{N_q}
\in\mathbb{R}^{N_q\times C}$. Geometry observations identify the field to be
recovered, whereas query coordinates specify its evaluation locations. This
task admits two distinct formulations according to the role of the query set.

In a set-to-set formulation, the geometry observation and the query set jointly
determine a finite output:
\begin{equation}
    \widehat{\mathbf{Y}}_{Q}
    =
    \mathcal{M}_{\theta}(X,Q),
    \qquad
    \left[\widehat{\mathbf{Y}}_{Q}\right]_{j}
    =
    f_{\theta}(X,\mathbf{q}_{j};Q).
    \label{eq:set-to-set-map}
\end{equation}
The subscript $Q$ emphasizes that the output is constructed for the current
query set. Attention, message passing, or set aggregation between query points
can therefore contribute directly to each prediction. The resulting map
approximates a joint discrete response on $Q$, and the value at a shared
coordinate can vary with the surrounding query context.

In a set-to-function formulation, the geometry observation first determines a
field representation, after which coordinates act as independent evaluation
variables:
\begin{equation}
    \mathbf{z}
    =
    \mathcal{E}_{\theta}(X),
    \qquad
    \widehat{\mathbf{y}}_{X}(\mathbf{q})
    =
    \mathcal{F}_{\theta}(\mathbf{z},\mathbf{q}),
    \qquad
    \widehat{\mathbf{Y}}(Q)
    =
    \left[
        \widehat{\mathbf{y}}_{X}(\mathbf{q}_{j})
    \right]_{j=1}^{N_q}.
    \label{eq:set-to-function-factorization}
\end{equation}
Here, $\mathbf{z}$ is constructed before output queries are introduced, and
$\widehat{\mathbf{y}}_{X}$ is the geometry-conditioned field associated with
that state. The query set selects evaluation coordinates but does not enter the
construction of either $\mathbf{z}$ or the field representation. When
coordinate features and field values are evaluated independently across
queries, any nested query sets $Q_1\subseteq Q_2$ satisfy
\begin{equation}
    \mathcal{R}_{Q_1}
    \left[
        \widehat{\mathbf{y}}_{X}(\mathbf{q})
    \right]_{\mathbf{q}\in Q_2}
    =
    \left[
        \widehat{\mathbf{y}}_{X}(\mathbf{q})
    \right]_{\mathbf{q}\in Q_1},
    \label{eq:query-restriction-consistency}
\end{equation}
where $\mathcal{R}_{Q_1}$ restricts the output on $Q_2$ to the shared
coordinates in $Q_1$. The same geometry state can consequently be reused
across resolutions, local regions, and query batches.

The two formulations approximate different numerical objects. A set-to-set map
constructs a discrete solution jointly on the current point set, whereas a
set-to-function map parameterizes a geometry-conditioned field before applying
a finite observation. The latter separates field construction from output
discretization. Its numerical realization must recover the global field
organization governed by the geometry from finite observations while supplying
coordinate-wise approximation capacity commensurate with local field
complexity.

\subsection{Architecture overview}
\label{sec:method-overview}

\begin{figure}[t]
    \centering
    \includegraphics[width=\textwidth]{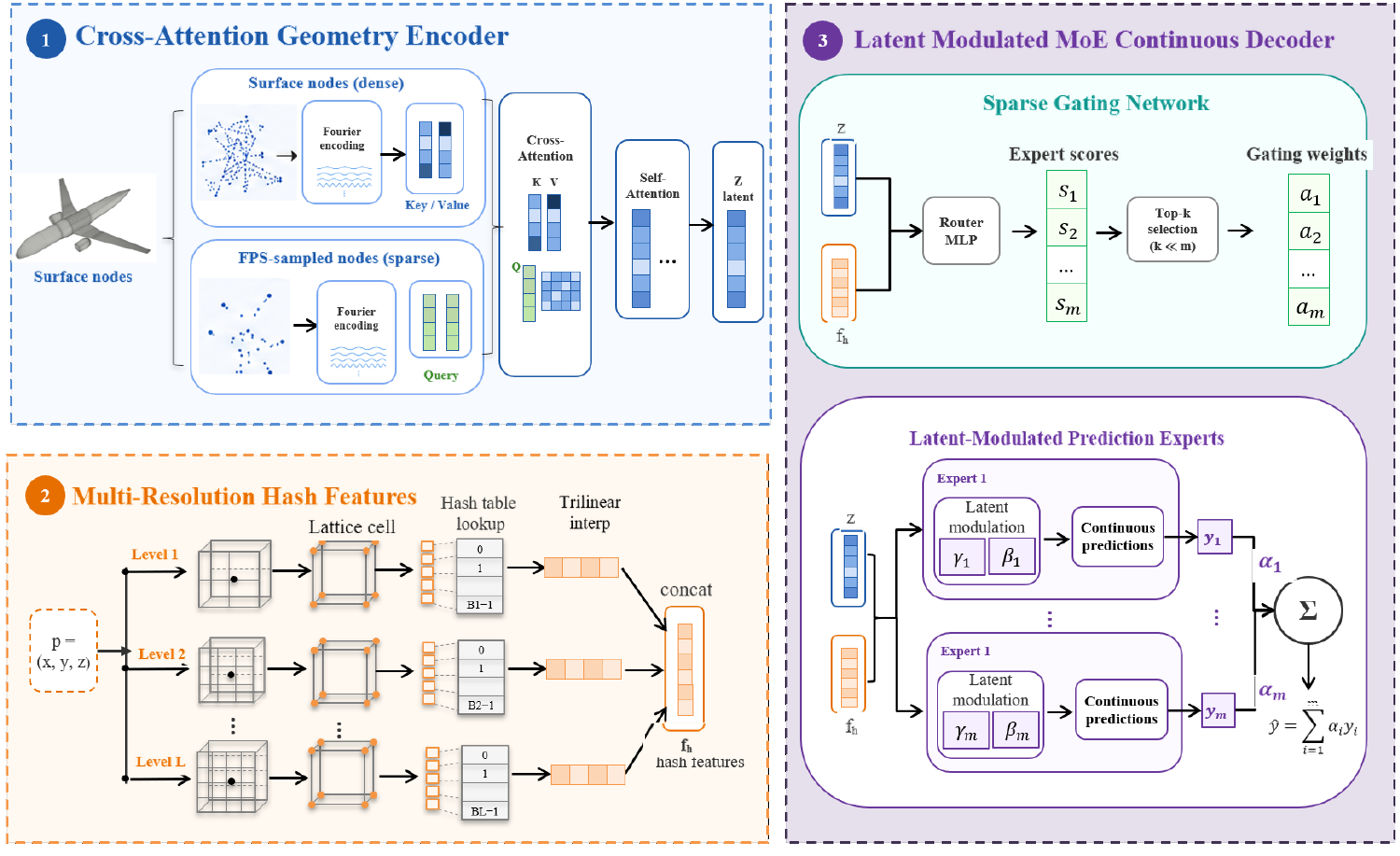}
    \caption{Architecture of GeoCANO. A point-set encoder constructs a global
    geometry state that modulates a multiresolution axis-factorized coordinate
    representation and conditions query-wise sparse expert decoding.}
    \label{fig:method-overview}
\end{figure}

Figure~\ref{fig:method-overview} summarizes the GeoCANO architecture and its
geometry-conditioned information flow.

For an aircraft configuration $g$, let $\Gamma_g\subset\mathbb{R}^{3}$ denote
the physical surface. A common affine transformation $\mathcal{T}_{x}$,
determined from the training configurations, maps the surface to
$\overline{\Gamma}_g=\mathcal{T}_{x}(\Gamma_g)\subset[0,1]^3$. The network
input is the unordered coordinate set
$\mathbf{X}_g^{(N_e)}
=\{\mathcal{T}_{x}(\mathbf{x}_{g,i})\}_{i=1}^{N_e}
\subset\overline{\Gamma}_g$.
The physical outputs are collected into the surface-state vector
$\mathbf{s}_g(\mathbf{x})
=[C_{p,g}(\mathbf{x}),\rho_g(\mathbf{x}),T_g(\mathbf{x})]^{\mathsf T}$.
With the channel-wise transformation $\mathcal{T}_{y}$ determined from the
training data, the normalized prediction target is
\begin{equation}
    \widetilde{\mathbf{y}}_g(\mathbf{q})
    =
    \mathcal{T}_{y}
    \!\left(
        \mathbf{s}_g
        \!\left(\mathcal{T}_{x}^{-1}(\mathbf{q})\right)
    \right)
    \in\mathbb{R}^{3},
    \qquad
    \mathbf{q}\in\overline{\Gamma}_g.
    \label{eq:normalized-surface-state}
\end{equation}

GeoCANO realizes the set-to-function construction through a global geometry
stage followed by coordinate-wise field evaluation. The point-set encoder first compresses
$\mathbf{X}_g^{(N_e)}$ into $M$ latent geometry tokens and forms a global
geometry state by mean pooling
\begin{equation}
    \mathbf{Z}_g
    =
    \mathcal{E}_{\theta_{\mathrm{enc}}}
    \!\left(\mathbf{X}_g^{(N_e)}\right)
    \in\mathbb{R}^{M\times d_z},
    \qquad
    \mathbf{z}_g
    =
    \frac{1}{M}
    \sum_{m=1}^{M}\mathbf{Z}_{g,m}
    \in\mathbb{R}^{d_z}.
    \label{eq:geometry-latent}
\end{equation}
The token sequence $\mathbf{Z}_g$ retains spatially distributed geometric
information, while $\mathbf{z}_g$ provides the configuration-level condition
shared by the coordinate representation and the decoder.

The global geometry state modulates the rank contributions of a shared
multiresolution axis-factorized representation. At a query coordinate
$\mathbf{q}_{g,j}$, the resulting feature is written compactly as
$\boldsymbol{\phi}_{g,j}
=\Phi_{\theta_{\mathrm{coord}}}
(\mathbf{q}_{g,j};\mathbf{z}_g)
\in\mathbb{R}^{d_{\phi}}$, with $d_{\phi}=3+LF$. It combines the original
coordinate with features from $L$ spatial resolutions. Concatenation with the
global geometry state gives the shared query descriptor
\begin{equation}
    \boldsymbol{\xi}_{g,j}
    =
    \begin{bmatrix}
        \mathbf{z}_g\\
        \boldsymbol{\phi}_{g,j}
    \end{bmatrix}
    \in\mathbb{R}^{d_z+d_{\phi}}.
    \label{eq:shared-query-descriptor}
\end{equation}
This descriptor couples configuration-level field organization with the local
coordinate representation required for field evaluation.

The router maps $\boldsymbol{\xi}_{g,j}$ to $E$ expert scores and retains the
$k_{\mathrm{act}}$ largest entries, forming the active set
$\mathcal{A}_{g,j}$. A softmax over this set gives the normalized weights
$\pi_{g,j,e}$. Each active expert produces a candidate state
$\widehat{\widetilde{\mathbf{y}}}_{g,j}^{(e)}
=\mathcal{D}_{\theta_{\mathrm{dec},e}}
(\boldsymbol{\xi}_{g,j};\mathbf{z}_g)\in\mathbb{R}^{3}$, and the final
prediction is
\begin{equation}
    \widehat{\widetilde{\mathbf{y}}}_g
    \!\left(\mathbf{q}_{g,j}\right)
    =
    \sum_{e\in\mathcal{A}_{g,j}}
    \pi_{g,j,e}\,
    \widehat{\widetilde{\mathbf{y}}}_{g,j}^{(e)}.
    \label{eq:end-to-end-map}
\end{equation}

The architecture therefore separates configuration-level field construction
from query-level realization. The geometry-dependent quantities
$\mathbf{Z}_g$, $\mathbf{z}_g$, and the associated coordinate modulation are
computed once for each configuration and reused across query sets. 

\subsection{Point set geometry encoder}
\label{sec:geometry-encoder}

The geometry encoder compresses an unordered surface-point set into a fixed
number of latent tokens that characterize the aircraft configuration. This
compression uses only point coordinates and does not require mesh connectivity.
For each configuration, $N_e$ encoding points are sampled from the archived
surface, after which farthest-point sampling (FPS) selects $M\ll N_e$
geometric anchors.

Before attention, each normalized coordinate is concatenated with its sine and
cosine components at the dyadic frequencies $2^k$, $k=0,\ldots,K_f-1$. This
pointwise Fourier lifting, denoted by $\boldsymbol\psi$, retains the original
coordinate while augmenting it with multiscale positional information. The FPS
anchors and the corresponding pointwise embeddings are
\begin{equation}
    \mathbf{A}_g
    =\operatorname{FPS}_{M}\!\left(\mathbf{X}_g^{(N_e)}\right),
    \qquad
    \mathbf{H}_g
    =\mathcal{P}_{\mathrm{in}}\!\left(
    \boldsymbol\psi(\mathbf{X}_g^{(N_e)})\right),
    \qquad
    \mathbf{H}_g^{A}
    =\mathcal{P}_{\mathrm{in}}\!\left(
    \boldsymbol\psi(\mathbf{A}_g)\right),
    \label{eq:point-anchor-embedding}
\end{equation}
where $\boldsymbol\psi$ is applied pointwise and
$\mathcal{P}_{\mathrm{in}}$ is a shared affine projection to width $d_a$.

The anchor embeddings serve as query tokens and aggregate information from all
surface tokens through a pre-normalized multihead cross-attention block,
\begin{equation}
\begin{aligned}
    \widetilde{\mathbf{U}}_g^{(0)}
    &=\mathbf{H}_g^{A}
    +\operatorname{MHA}_{c}\!\left(
        \operatorname{LN}(\mathbf{H}_g^{A}),
        \operatorname{LN}(\mathbf{H}_g),
        \operatorname{LN}(\mathbf{H}_g)
    \right),\\
    \mathbf{U}_g^{(0)}
    &=\widetilde{\mathbf{U}}_g^{(0)}
    +\operatorname{FFN}_{c}\!\left(
        \operatorname{LN}(\widetilde{\mathbf{U}}_g^{(0)})
    \right).
\end{aligned}
\label{eq:cross-attention-compression}
\end{equation}
Here, $\operatorname{MHA}_{c}$ denotes scaled dot-product multihead
cross-attention and $\operatorname{FFN}_{c}$ is a two-layer GELU feed-forward
map. The compressed anchor sequence is subsequently refined by $S$
pre-normalized self-attention blocks,
\begin{equation}
\begin{aligned}
    \widetilde{\mathbf{U}}_g^{(s)}
    &=\mathbf{U}_g^{(s-1)}
    +\operatorname{MHA}_{s}\!\left(
        \operatorname{LN}(\mathbf{U}_g^{(s-1)}),
        \operatorname{LN}(\mathbf{U}_g^{(s-1)}),
        \operatorname{LN}(\mathbf{U}_g^{(s-1)})
    \right),\\
    \mathbf{U}_g^{(s)}
    &=\widetilde{\mathbf{U}}_g^{(s)}
    +\operatorname{FFN}_{s}\!\left(
        \operatorname{LN}(\widetilde{\mathbf{U}}_g^{(s)})
    \right),
    \qquad s=1,\ldots,S,\\
    \mathbf{Z}_g
    &=\mathcal{P}_{z}\!\left(
        \operatorname{LN}(\mathbf{U}_g^{(S)})
    \right).
\end{aligned}
\label{eq:self-attention-refinement}
\end{equation}
Because $M\ll N_e$, the cross-attention stage compresses the full point set
with cost $\mathcal{O}(MN_ed_a)$ instead of applying quadratic self-attention
over all $N_e$ points. The subsequent self-attention stack operates only on the
$M$ anchor tokens and scales as $\mathcal{O}(SM^2d_a)$. Thus, $M$ controls both
the geometric representation budget and the encoder cost, rather than an
output-mesh resolution.

\subsection{Geometry-modulated multiresolution CP encoding}
\label{sec:mcp-encoding}
Direct coordinates require the decoder to learn multiscale spatial variation
solely through its hidden layers. GeoCANO therefore augments each query with a
shared multiresolution coordinate representation that combines one-dimensional
hash indexing, axis factorization, and geometry-dependent rank modulation. At
every level, trainable factors are queried independently along the three
aircraft-aligned Cartesian axes and coupled through common rank components. The
multiplicative construction is CP-structured, but it is learned directly rather
than obtained by decomposing a pre-existing dense tensor.

The encoder contains $L$ levels with resolutions
\begin{equation}
    n_\ell=\left\lfloor n_{\min}b^\ell\right\rfloor,
    \qquad
    b=\left(\frac{n_{\max}}{n_{\min}}\right)^{1/(L-1)},
    \qquad
    \ell=0,\ldots,L-1.
    \label{eq:hash-resolution}
\end{equation}
At level $\ell$, the model stores a single table
$\mathbf{H}_{\ell}\in\mathbb{R}^{3n_\ell\times R\times F}$. The three axes occupy
disjoint table partitions with offsets $o_\ell^{(x)}=0$,
$o_\ell^{(y)}=n_\ell$, and $o_\ell^{(z)}=2n_\ell$, thereby retaining separate
axis-specific factors within the shared table. A logical index $i$ is mapped to
a physical table index by
\begin{equation}
    h_\ell^{(a)}(i)
    =\operatorname{mod}\!\left(
        (ip_a)\operatorname{xor}\bigl((ip_a)\mathbin{\gg}16\bigr),
        n_\ell
    \right),
    \qquad
    \eta_\ell^{(a)}(i)=h_\ell^{(a)}(i)+o_\ell^{(a)},
    \label{eq:axis-hash-index}
\end{equation}
where $p_a$ is a fixed axis-specific prime. Because $h_\ell^{(a)}$ is not
required to preserve the original grid ordering, it provides a deterministic
multiresolution indexing scheme for the axis factors. Each retrieved row is
interpreted as $R$ coupled feature vectors of width $F$.

For $q_a\in[0,1]$, the level coordinate is $s_\ell(q_a)=n_\ell q_a$. The two
lookup indices are
$i_\ell^0(q_a)=\min\{\lfloor s_\ell(q_a)\rfloor,n_\ell-1\}$ and
$i_\ell^1(q_a)=\min\{i_\ell^0(q_a)+1,n_\ell-1\}$, with interpolation weight
$\lambda_\ell(q_a)=s_\ell(q_a)-\lfloor s_\ell(q_a)\rfloor$. The interpolated
feature of rank component $r$ along axis $a$ is
\begin{equation}
    \mathbf{f}_{\ell r}^{(a)}(q_a)
    =\left(1-\lambda_\ell(q_a)\right)
    \mathbf{H}_{\ell,\eta_\ell^{(a)}(i_\ell^0(q_a)),r}
    +\lambda_\ell(q_a)
    \mathbf{H}_{\ell,\eta_\ell^{(a)}(i_\ell^1(q_a)),r}
    \in\mathbb{R}^{F}.
    \label{eq:axis-hash-interpolation}
\end{equation}

Geometry dependence is introduced by residual reweighting of the coupled rank
components. A two-layer modulation network $\mathcal{M}_{\alpha}$ gives
\begin{equation}
    \Delta\boldsymbol\alpha_g
    =\mathcal{M}_{\alpha}(\mathbf{z}_g)
    \in\mathbb{R}^{L\times R},
    \qquad
    \boldsymbol\alpha_g
    =\mathbf{1}_{L\times R}
    +\Delta\boldsymbol\alpha_g.
    \label{eq:rank-modulation}
\end{equation}
The final affine layer of $\mathcal{M}_{\alpha}$ is initialized to zero, making
the initial encoder equivalent to the shared unmodulated representation. At
query $\mathbf{q}_{g,j}$, the level-$\ell$ feature is
\begin{equation}
    \boldsymbol\phi_{g,j}^{(\ell)}
    =\sum_{r=1}^{R}\alpha_{g,\ell r}
    \left[
        \mathbf{f}_{\ell r}^{(x)}(q_{g,j,x})
        \odot
        \mathbf{f}_{\ell r}^{(y)}(q_{g,j,y})
        \odot
        \mathbf{f}_{\ell r}^{(z)}(q_{g,j,z})
    \right]
    \in\mathbb{R}^{F},
    \label{eq:factorized-level-feature}
\end{equation}
and the complete coordinate representation is
\begin{equation}
    \boldsymbol\phi_{g,j}
    =\operatorname{concat}\!\left(
        \mathbf{q}_{g,j},
        \boldsymbol\phi_{g,j}^{(0)},\ldots,
        \boldsymbol\phi_{g,j}^{(L-1)}
    \right)
    \in\mathbb{R}^{3+LF}.
    \label{eq:multiresolution-query-feature}
\end{equation}
The query coordinate determines the one-dimensional hash lookup and
interpolation along each axis, whereas $\boldsymbol\alpha_g$ globally adjusts
the rank contributions for the complete configuration and is shared across its
query locations. The axis-partitioned lookup supplies multiresolution spatial
features, while the CP-structured product provides rank-controlled cross-axis
interaction. Because the factors are aligned with the Cartesian axes, all
configurations are expressed in the same aircraft-fixed frame and normalized by
the same training-set-derived transformation.

Conventional multiresolution three-dimensional hash encoding first locates a
query in the grid associated with each resolution level. The eight vertices of
the containing cell are mapped to a finite feature table, and their retrieved
features are combined by trilinear interpolation.  GeoCANO instead
performs one-dimensional lookup along each axis and reconstructs the cross-axis
interaction through a rank-$R$ CP structure. The resulting factorization avoids
maintaining a large joint three-dimensional table, while the latent-dependent
rank weights adapt the shared multiresolution factors to each aircraft
configuration. The parameter counts of the two coordinate encoders are
\begin{equation}
    P_{\mathrm{factorized}}
    =3RF\sum_{\ell=0}^{L-1}n_\ell
    +H_m(d_z+1)+LR(H_m+1),
    \qquad
    P_{\mathrm{3D\text{-}hash}}
    =F\sum_{\ell=0}^{L-1}T_\ell^{(3D)},
    \label{eq:hash-parameter-count}
\end{equation}
respectively, where $H_m$ is the hidden width of $\mathcal{M}_{\alpha}$ and
$T_\ell^{(3D)}=\min\{n_\ell^3,T_{\max}\}$ is the allocated table capacity of the
conventional three-dimensional hash encoder at level $\ell$. The dominant
storage term of the proposed representation scales with
$3R\sum_\ell n_\ell$, whereas that of the conventional encoder is determined by
$\sum_\ell T_\ell^{(3D)}$. The factorized construction therefore provides a
more compact coordinate representation when the three-dimensional hash tables
require substantially larger level-wise capacities. Linear interpolation
permits direct feature evaluation at any admissible normalized query coordinate
without first predicting on a fixed output mesh.

\subsection{Query-wise sparse mixture-of-experts decoder}
\label{sec:moe-decoder}

The query-wise sparse decoder allocates prediction capacity according to both
the aircraft configuration and the requested surface location. All experts are
defined over the shared descriptor in
Eq.~\eqref{eq:shared-query-descriptor}, but only the experts selected for a
given query are evaluated. Their affine transformations, geometry-modulation
networks, and readout heads are parameterized independently.

For expert $e\in\{1,\ldots,E\}$, a modulation network generates one
scale--shift pair for each of $K$ hidden blocks of width $d_h$,
\begin{equation}
    \left\{\mathbf{s}_{g,e}^{(k)},\mathbf{t}_{g,e}^{(k)}\right\}_{k=1}^{K}
    =\mathcal{M}_{\theta_{M,e}}(\mathbf{z}_g),
    \qquad
    \mathbf{s}_{g,e}^{(k)},\mathbf{t}_{g,e}^{(k)}\in\mathbb{R}^{d_h}.
    \label{eq:expert-modulation}
\end{equation}
With $\mathbf{h}_{g,j,e}^{(0)}=\mathbf{0}$, the hidden state is updated as
\begin{equation}
    \mathbf{h}_{g,j,e}^{(k)}
    =\operatorname{GELU}\!\left(
        \mathbf{s}_{g,e}^{(k)}\odot
        \left(\mathbf{W}_{e}^{(k)}\boldsymbol\xi_{g,j}
        +\mathbf{b}_{e}^{(k)}\right)
        +\mathbf{t}_{g,e}^{(k)}
        +\mathbf{h}_{g,j,e}^{(k-1)}
    \right),
    \qquad k=1,\ldots,K.
    \label{eq:additive-expert-recurrence}
\end{equation}
Geometry therefore enters each expert through two complementary paths. The
descriptor $\boldsymbol\xi_{g,j}$ directly supplies the global geometry code
and the local coordinate representation to every hidden block, whereas
$\mathbf{z}_g$ generates feature-wise scales and shifts that adapt the block
response to the aircraft configuration. Channel-specific heads map the final
hidden state to the normalized physical variables,
\begin{equation}
    \widehat{\widetilde y}_{g,j,c}^{(e)}
    =\mathcal{H}_{\theta_{H,e,c}}
    \!\left(\mathbf{h}_{g,j,e}^{(K)}\right),
    \qquad c=1,\ldots,C,
    \label{eq:expert-field-heads}
\end{equation}
and the corresponding expert prediction is denoted by
$\widehat{\widetilde{\mathbf{y}}}_{g,j}^{(e)}\in\mathbb{R}^{C}$.

The router maps the same descriptor to $E$ logits and retains the
$k_{\mathrm{act}}$ largest entries,
\begin{equation}
    \mathbf{r}_{g,j}
    =\tau^{-1}\mathcal{R}_{\theta_R}(\boldsymbol\xi_{g,j}),
    \qquad
    \mathcal{A}_{g,j}
    =\operatorname{TopK}(\mathbf{r}_{g,j},k_{\mathrm{act}}),
    \label{eq:router-selection}
\end{equation}
where $\tau>0$ is the fixed routing temperature. The normalized weight of an
active expert is
\begin{equation}
    \pi_{g,j,e}
    =\frac{\exp(r_{g,j,e})}
    {\sum_{m\in\mathcal{A}_{g,j}}\exp(r_{g,j,m})},
    \qquad e\in\mathcal{A}_{g,j},
    \label{eq:sparse-routing-weights}
\end{equation}
with zero weight assigned to inactive experts. The final mixture prediction is
\begin{equation}
    \widehat{\widetilde{\mathbf{y}}}_{g,j}
    =\sum_{e\in\mathcal{A}_{g,j}}
    \pi_{g,j,e}\widehat{\widetilde{\mathbf{y}}}_{g,j}^{(e)}.
    \label{eq:moe-output}
\end{equation}
The adopted configuration uses top-2 routing among four experts. No
expert-capacity constraint is imposed, and no query is dropped after the
top-$k$ assignment. Because the descriptor can be constructed at any
admissible coordinate, Eq.~\eqref{eq:moe-output} can be evaluated directly at
arbitrary query locations. This queryability does not imply global smoothness:
the active set can change at logit-ordering boundaries.

\subsection{Training objective and optimization}
\label{sec:training-objective}

The normalized mixture prediction in Eq.~\eqref{eq:moe-output} is trained with
a channel-weighted mean-squared error. For a minibatch of $B$ configurations,
$N_q$ queries per configuration, and $C$ output channels, the data term is
\begin{equation}
    \mathcal{L}_{\mathrm{data}}
    =\frac{1}{BN_qC}
    \sum_{g=1}^{B}\sum_{j=1}^{N_q}\sum_{c=1}^{C}
    w_c
    \left(
        \widehat{\widetilde y}_{g,j,c}
        -\widetilde y_{g,j,c}
    \right)^2,
    \label{eq:data-loss}
\end{equation}
where the positive channel weights $w_c$ are determined once from training-set
statistics before optimization and subsequently held fixed. Neither the field
normalization statistics nor the channel weights are recomputed on the test
set.

To discourage highly imbalanced expert utilization, the routing regularization
combines the dense routing probabilities with the realized top-$k$
assignments. Flattening the batch and query indices into $n=1,\ldots,N$, where
$N=BN_q$, define
\begin{equation}
    p_{n,e}=\operatorname{softmax}(\mathbf{r}_n)_e,
    \qquad
    I_e=\frac{1}{N}\sum_{n=1}^{N}p_{n,e},
    \qquad
    F_e=\frac{1}{N}\sum_{n=1}^{N}
    \mathbb{I}(e\in\mathcal{A}_n).
    \label{eq:router-importance-load}
\end{equation}
Here, $I_e$ is the mean dense routing probability and $F_e$ is the fraction of
queries assigned to expert $e$. The hard assignment frequency $F_e$ is treated
as stop-gradient when forming the balancing term. The routing regularizers are
\begin{equation}
\begin{aligned}
    \mathcal{L}_{\mathrm{bal}}
    &=E\sum_{e=1}^{E}I_eF_e,\\
    \mathcal{L}_{z}
    &=\frac{1}{N}\sum_{n=1}^{N}
    \left[
        \log\!\left(\sum_{e=1}^{E}\exp(r_{n,e})\right)
    \right]^2,
\end{aligned}
\label{eq:router-regularizers}
\end{equation}
The load-balancing term favors a more even distribution of assignments, while
the router $z$-loss limits excessive logit growth. Because each query activates
$k_{\mathrm{act}}$ experts, $\sum_eF_e=k_{\mathrm{act}}$; the resulting constant
scale is absorbed into the regularization weight. The complete objective is
\begin{equation}
    \mathcal{L}
    =\mathcal{L}_{\mathrm{data}}
    +\lambda_{\mathrm{bal}}\mathcal{L}_{\mathrm{bal}}
    +\lambda_z\mathcal{L}_{z}.
    \label{eq:total-loss}
\end{equation}

\subsection{Aerodynamic performance prediction}
\label{sec:load-reconstruction}

Although connectivity and surface metrics are excluded from the neural input, they can be reintroduced after inference to evaluate whether the recovered pressure field preserves integrated aerodynamic behavior. Let $\mathcal{T}_g$ denote the set of surface faces for configuration $g$. For face $f\in\mathcal{T}_g$, let $\mathcal{V}_{g,f}$ be its vertex-index set, $A_{g,f}$ its physical area, $\mathbf{n}_{g,f}$ its outward unit normal, and $\mathbf{x}^{\mathrm{phys}}_{g,f}$ its physical centroid. After applying the inverse field normalization, the face-centered predicted pressure coefficient is obtained from the nodal predictions as
\begin{equation}
    \widehat{\overline C}_{p,g,f}
    =\frac{1}{|\mathcal{V}_{g,f}|}
    \sum_{i\in\mathcal{V}_{g,f}}
    \widehat C_{p,g}(\mathbf{q}_{g,i}),
    \label{eq:face-pressure-quadrature}
\end{equation}
For the low-order surface elements used in the archived CFD meshes, the corresponding pressure-force and pressure-moment coefficients are approximated by
\begin{equation}
\begin{aligned}
    \widehat{\mathbf{C}}_{F,p,g}
    &=-\frac{1}{S_{\mathrm{ref}}}
    \sum_{f\in\mathcal{T}_g}
    \widehat{\overline C}_{p,g,f}
    A_{g,f}\mathbf{n}_{g,f},\\
    \widehat{\mathbf{C}}_{M,p,g}
    &=-\frac{1}{S_{\mathrm{ref}}L_{\mathrm{ref}}}
    \sum_{f\in\mathcal{T}_g}
    (\mathbf{x}^{\mathrm{phys}}_{g,f}-\mathbf{x}^{\mathrm{phys}}_{\mathrm{ref}})
    \times
    \left(
    \widehat{\overline C}_{p,g,f}
    A_{g,f}\mathbf{n}_{g,f}
    \right),
\end{aligned}
\label{eq:pressure-load-quadrature}
\end{equation}
where $\mathbf{q}_{g,i}$ is the normalized network query corresponding to physical surface vertex $i$, and $S_{\mathrm{ref}}$, $L_{\mathrm{ref}}$, and $\mathbf{x}^{\mathrm{phys}}_{\mathrm{ref}}$ are the aerodynamic reference area, reference length, and moment center used by the CFD database. The pressure contributions to lift and drag follow by projection onto the corresponding unit directions,
\begin{equation}
    \widehat C_{L,p,g}
    =\mathbf{e}_{L,g}^{\!\top}\widehat{\mathbf{C}}_{F,p,g},
    \qquad
    \widehat C_{D,p,g}
    =\mathbf{e}_{D,g}^{\!\top}\widehat{\mathbf{C}}_{F,p,g}.
    \label{eq:pressure-lift-drag}
\end{equation}
Any symmetry convention used by the half-model CFD mesh is applied identically to the predicted and reference quadratures. Equations~\eqref{eq:face-pressure-quadrature}--\eqref{eq:pressure-lift-drag} are not part of the learning map or the training loss. They provide a deterministic, engineering-level diagnostic that propagates local $C_p$ errors into global pressure-load discrepancies without introducing mesh information into the network.

\section{Dataset construction}
\label{sec:dataset}

\paragraph{\textbf{Geometry parameterization and sampling.}}
\label{sec:geometry_sampling}

The design family $\mathcal{G}$ comprises wing--body--nacelle geometries
representative of a narrow-body transport-aircraft class. The fuselage outer
mold line and the nacelle body geometry remain fixed throughout the family,
whereas the wing planform, spanwise airfoil assignment, and geometric twist
are varied. Although the nacelle outer mold line is invariant, its installation
pose is updated with the local wing geometry to preserve the prescribed
relative mounting arrangement and prevent component intersections. This
construction allows configuration-dependent wing--body and wing--nacelle
interaction effects to be represented within a controlled family of wing-shape
variations.

\begin{figure}[htpb]
    \centering
    \includegraphics[width=0.95\linewidth]{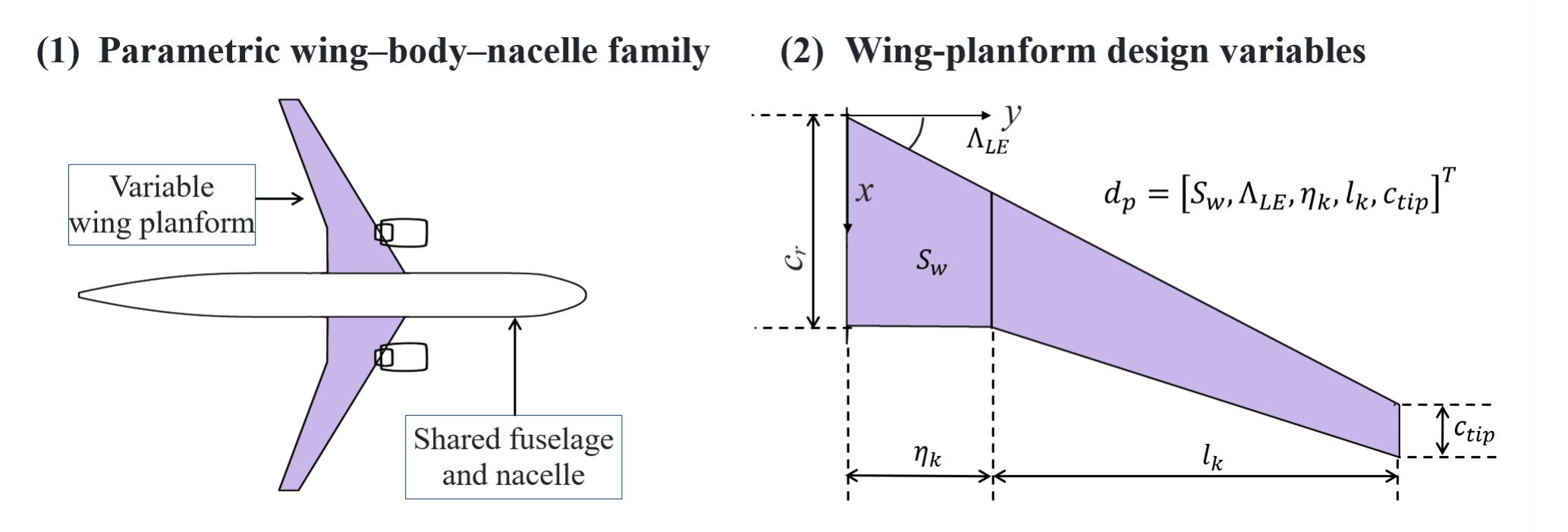}
    \caption{Controlled parameterization of the wing--body--nacelle geometry
    family. The fuselage outer mold line and nacelle body geometry are fixed,
    whereas the wing is varied at eight spanwise control stations.}
    \label{fig:geometry-parameterization}
\end{figure}

Each aircraft configuration is parameterized by a planform descriptor
$\mathbf{d}_p$ and an ordered spanwise-section descriptor $\mathbf{d}_s$:
\begin{equation}
    \mathbf{d}=(\mathbf{d}_p,\mathbf{d}_s),
    \qquad
    \mathbf{d}_p
    =\left(S_w,\Lambda_{\mathrm{LE}},\eta_k,\ell_k,c_{\mathrm{tip}}\right),
    \qquad
    \mathbf{d}_s
    =\bigl((a_r,\tau_r)\bigr)_{r=1}^{8}.
    \label{eq:geometry-design-vector}
\end{equation}
Here, $S_w$ denotes the wing semispan, $\Lambda_{\mathrm{LE}}$ the
leading-edge sweep angle, $\eta_k$ the spanwise coordinate of the planform
kink, $\ell_k$ the parameter governing the extent of the outboard wing
segment, and $c_{\mathrm{tip}}$ the tip chord. At control station $r$, $a_r$
identifies the assigned airfoil template and $\tau_r$ denotes the local
geometric twist. Equation~\eqref{eq:geometry-design-vector} specifies the CAD
parameters used to generate the geometry library. These descriptors are not
provided to the neural operator; its geometric input consists only of a sampled
set of surface coordinates.
Figure~\ref{fig:geometry-parameterization} summarizes the controlled
parameterization used to generate the geometry family.

Controlled perturbations were applied to the wing planform, spanwise airfoil
assignment, and twist distribution within their prescribed parameter ranges to
cover both individual variations and coupled changes in multiple quantities.
Before three-dimensional RANS calculations were undertaken, candidate spanwise
sections were screened using a low-cost aerodynamic analysis. Configurations
whose pressure-recovery behavior or shock-related features fell outside the
admissible range were locally revised by modifying the leading-edge radius,
camber, or thickness distribution; those that remained unacceptable were excluded.
This procedure produced $11{,}500$ nominal aircraft configurations.

\paragraph{\textbf{RANS simulations.}}
\label{sec:cfd_label_generation}

\begin{figure}[t]
    \centering
    \begin{subfigure}[t]{0.44\linewidth}
        \centering
        \includegraphics[width=\linewidth]{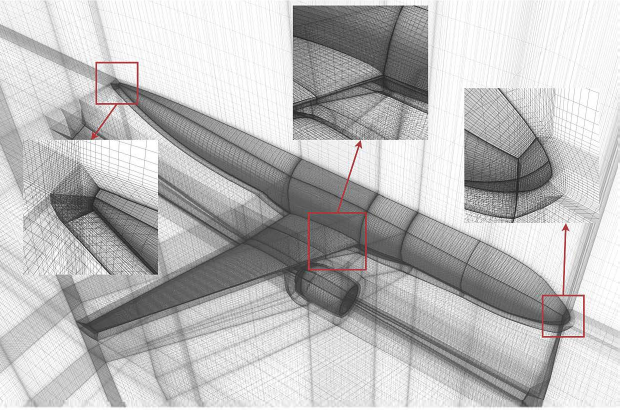}
        \caption{Multiblock volume grid with targeted local refinement.}
        \label{fig:representative-mesh}
    \end{subfigure}
    \hfill
    \begin{subfigure}[t]{0.54\linewidth}
        \centering
        \includegraphics[width=\linewidth]{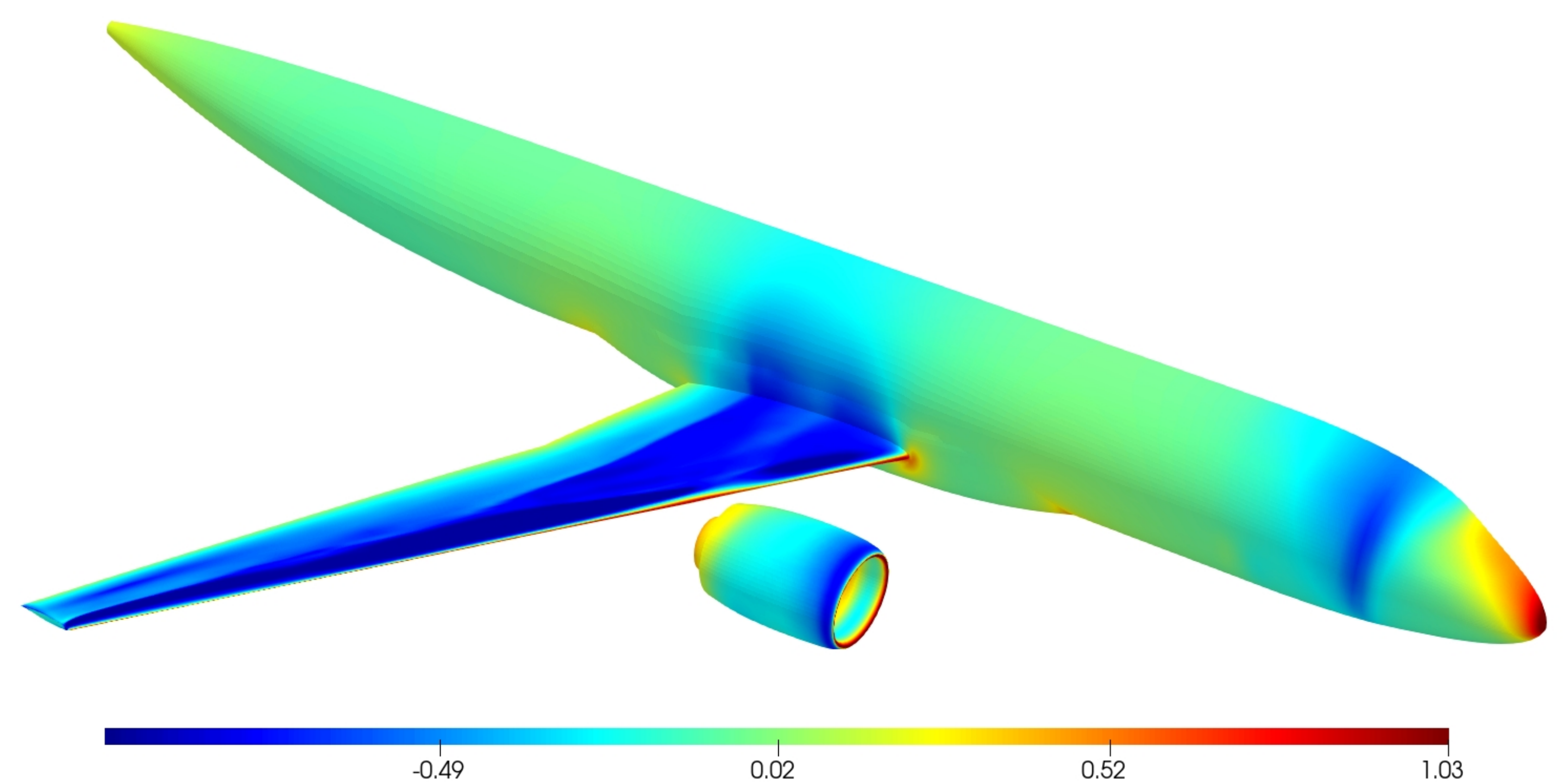}
        \caption{Surface pressure coefficient extracted from an accepted RANS solution.}
        \label{fig:representative-surface-field}
    \end{subfigure}
    \caption{Representative structured-grid discretization and surface-field
    extraction. The common multiblock topology and refinement strategy are
    applied throughout the geometry family to control discretization-induced
    case-to-case variability.}
    \label{fig:mesh-and-surface-field}
\end{figure}

To control numerical variability across the geometry family, each nominal
configuration satisfying the geometry-validity checks was discretized using a
common multiblock structured-grid strategy. In ICEM CFD, a predefined block
topology was adapted to the physical surface of each configuration, while the
block connectivity, near-wall construction, and region-specific refinement
rules were retained. The far field employed an H-type topology, and body-fitted
O-type blocks resolved the viscous region adjacent to the aircraft surface.
Each half-model grid contained approximately $12\times10^{6}$ cells. The first
wall-normal spacing targeted $y^{+}\approx1$, the wall-normal growth ratio was
1.2, and the outer boundary was placed at a distance of $10L_{\mathrm{ref}}$
from the aircraft, where $L_{\mathrm{ref}}$ denotes the reference length
adopted in the CFD setup. Additional refinement was applied near the wing
leading and trailing edges, the nacelle inlet, and the wing--body and
wing--nacelle junctions. Applying the same topology and resolution rules to all
configurations controlled case-to-case variation associated with changes in the
prescribed discretization, while allowing the grid coordinates to conform to
each geometry.
Figure~\ref{fig:mesh-and-surface-field} shows a representative grid and the
corresponding extracted surface pressure field.

All simulations followed a common operating and numerical protocol. The
free-stream Mach number was fixed at $M_\infty=0.78$, and a lift coefficient of
0.57 was prescribed; the boundary conditions and solver controls were applied
consistently throughout the campaign. For each configuration, the angle of
attack $\alpha_g$ was adjusted within the CFD procedure until the computed lift
coefficient reached the prescribed value. Accordingly, $\alpha_g$ was a
geometry-dependent control variable determined by the numerical protocol,
rather than an independent design variable or a network input.

CFL3D was used to solve the steady three-dimensional compressible
Reynolds-averaged Navier--Stokes equations. The system was closed with Menter's
shear-stress-transport $k$--$\omega$ turbulence model. Convective fluxes were
discretized using a second-order upwind Roe scheme, and the discrete steady
system was advanced by an implicit lower--upper symmetric Gauss--Seidel method.
Numerical acceptance was assessed jointly through residual reduction and
target-lift fulfillment. The residual measure was the $L_2$ norm of the
discrete RANS residual normalized by its value at the initial iterate. The
same convergence requirement was applied to every configuration, and a case was
retained only after the residual had decreased to the prescribed level and the computed lift coefficient had reached 0.57 to the adopted numerical accuracy.

For each numerically accepted configuration, the archived surface state
comprises the collocated values of $C_p$, $\rho$, and $T$ evaluated at all
$N_s=82{,}409$ surface nodes. The complete surface-coordinate and state
matrices are denoted by
\begin{equation}
    \begin{aligned}
        \mathbf{V}_g
        &=
        \begin{bmatrix}
            \mathbf{v}_{g,1} & \cdots & \mathbf{v}_{g,N_s}
        \end{bmatrix}^{\mathsf T}
        \in\mathbb{R}^{N_s\times3},\\
        \mathbf{Y}_g
        &=
        \begin{bmatrix}
            \mathbf{s}_g(\mathbf{v}_{g,1}) & \cdots &
            \mathbf{s}_g(\mathbf{v}_{g,N_s})
        \end{bmatrix}^{\mathsf T}
        \in\mathbb{R}^{N_s\times3}.
    \end{aligned}
    \label{eq:archived-surface-state}
\end{equation}
where $\mathbf{v}_{g,i}\in\Gamma_g$ and each row of $\mathbf{Y}_g$ contains
the collocated values of $C_p$, $\rho$, and $T$. Because the archived surface
meshes share a common topology and point correspondence, $N_s$ and the node
enumeration are consistent across configurations, although the physical
coordinates in $\mathbf{V}_g$ vary with $g$.

\paragraph{\textbf{Dataset curation and partition.}}
\label{sec:dataset_curation}

\begin{figure}[t]
    \centering
    \includegraphics[width=0.95\linewidth]{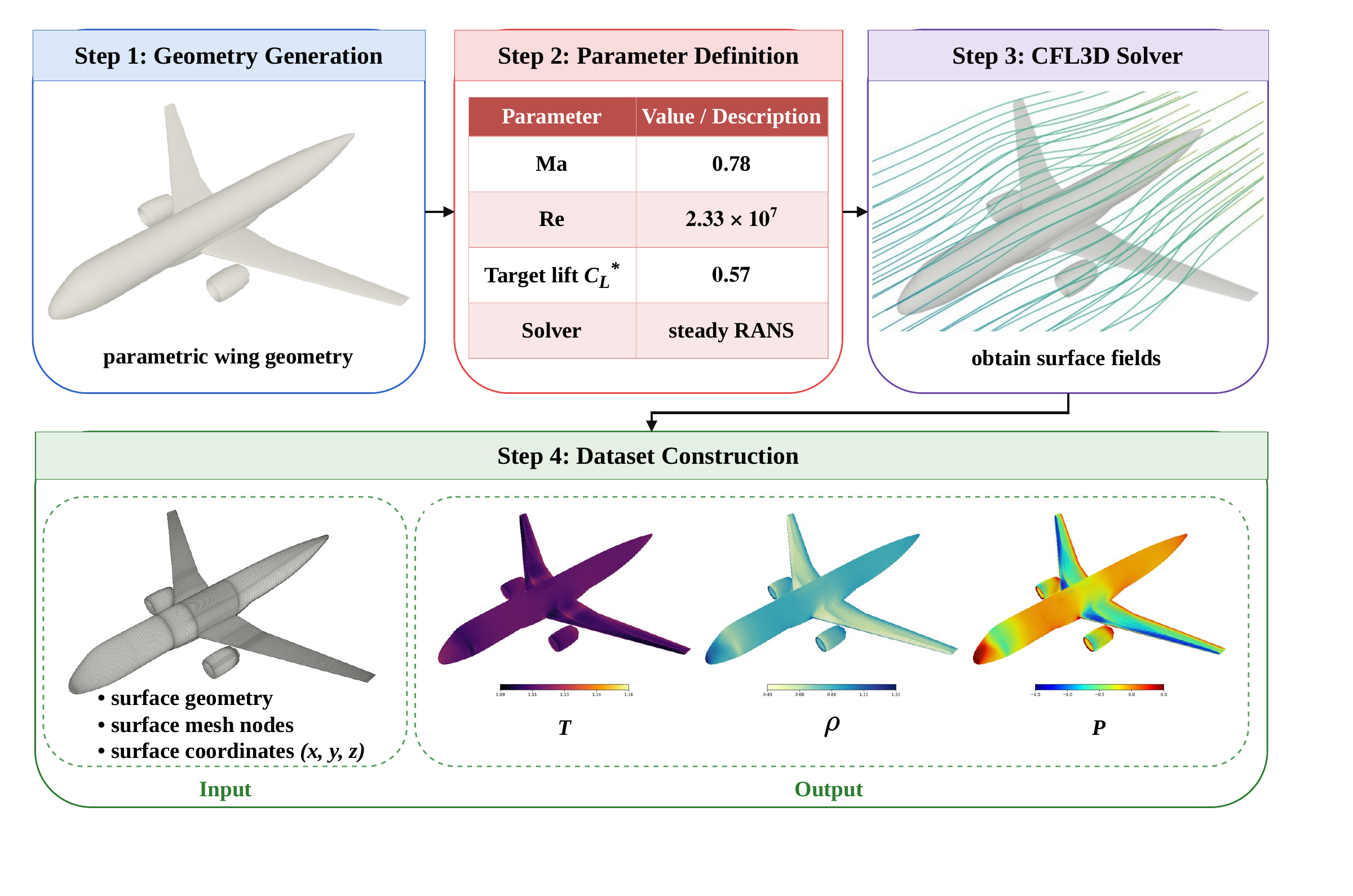}
    \caption{High-level construction of the geometry-conditioned aircraft
    surface-field dataset. Parametric wing--body--nacelle geometries are solved
    under a common target-lift RANS protocol, and accepted cases are archived as
    surface coordinates with collocated $C_p$, $\rho$, and $T$ labels. The four
    acceptance gates are applied before a nominal case enters the final
    database.}
    \label{fig:dataset-generation-pipeline}
\end{figure}

Of the $11{,}500$ nominal planform--section configurations, $10{,}484$ passed the four successive stages
of aerodynamic screening, CAD admissibility checks, mesh generation, and
numerical acceptance of the target-lift RANS solution. These accepted
geometry--solution pairs constitute the final CFD dataset.
Figure~\ref{fig:dataset-generation-pipeline} summarizes the workflow from
parametric geometry construction to the archived surface-coordinate and field
records.

For each accepted case, the archive retains the complete surface-coordinate
matrix $\mathbf{V}_g$ and state matrix $\mathbf{Y}_g$ defined in
Eq.~\eqref{eq:archived-surface-state}, together with the surface connectivity,
oriented face normals, face areas, pressure-integrated aerodynamic
coefficients, and case metadata. Within the learning pipeline, coordinates
sampled from $\mathbf{V}_g$ are mapped by $\mathcal{T}_x$ to form the normalized
geometry condition $\mathbf{X}_g^{(N_e)}$ and the field-query set
$\mathbf{Q}_g$. The corresponding CFD values are drawn from $\mathbf{Y}_g$ and
transformed by $\mathcal{T}_y$ to provide the supervision. Thus, only the
coordinate sets $\mathbf{X}_g^{(N_e)}$ and $\mathbf{Q}_g$ enter the neural
model; the remaining archived quantities are reserved for deterministic
post-processing, load integration, and evaluation.

The $10{,}484$ accepted cases were partitioned at the complete-configuration
level into a training set of $9{,}436$ cases and a test set of $1{,}048$ cases.
This partition was fixed before any geometry-encoding or field-query point
subsets were assembled. No coordinate sample, field label, or normalization
statistic derived from a test configuration was used for model fitting or
selection. The test results therefore quantify performance on complete
planform--section combinations unseen during training but contained within the
sampled design family.

\section{Numerical Results}
\label{sec:experiments}

\subsection{Experimental protocol and evaluation metrics}
\label{sec:experimental-setup}

The experiments used the 10,484 numerically accepted wing--body--nacelle configurations. Before any surface points were sampled, the complete configurations were partitioned at a ratio of $9{:}1$ into a training set and a test set. All samples associated with a given geometry inherited the assignment of that geometry. The reported results therefore quantify generalization to previously unseen configurations within the sampled design family. Given a sampled surface-coordinate set, the learned operator predicts the surface pressure coefficient $C_p$, density $\rho$, and temperature $T$.

\begin{figure}[!htbp]
    \centering
    \includegraphics[width=0.6\textwidth]{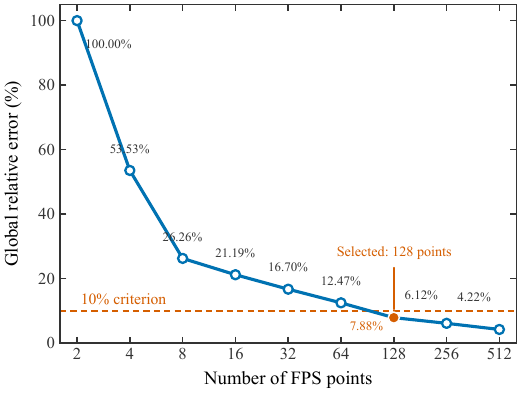}
    \caption{Relative deformation-reconstruction error versus the FPS anchor count. The dashed line marks the $10\%$ selection threshold, first satisfied at $M=128$.}
    \label{fig:fps-anchor-selection}
\end{figure}

For each training configuration, the geometry encoder received $N_e=10{,}000$ surface points. The encoder subset was fixed for each configuration, whereas the query subset was resampled between epochs to improve coverage of the aircraft surface. Each output channel was normalized independently using bounds determined exclusively from the training set. The encoder anchor budget $M$ was determined before model training through a deformation-reconstruction diagnostic based on the training geometries. Because the archived surface meshes share nodal correspondence, a reference surface was first obtained by averaging the coordinates at corresponding nodes over the training set, and each geometry was represented by its coordinate displacement from this reference. For each candidate value of $M$, FPS was applied to the reference surface, every surface node was assigned to its nearest anchor, and the displacement at that node was approximated by the displacement of the associated anchor. The relative reconstruction error was computed as the square root of the total squared deformation-reconstruction residual divided by the total deformation energy. A relative-error threshold of $10\%$, corresponding to a residual deformation-energy ratio below $1\%$, was used to screen the candidate anchor budgets. Figure~\ref{fig:fps-anchor-selection} shows that $M=128$ is the smallest tested value satisfying this criterion, with a relative deformation-reconstruction error of $7.88\%$. We therefore used 128 anchor tokens in all subsequent experiments, providing a fixed geometry-compression budget that is independent of the output-query resolution.

The model was optimized for 500 epochs using AdamW with an initial learning rate of $1.5\times10^{-3}$ and zero weight decay. A cosine learning-rate schedule was adopted with a batch size of 20 configurations, and the gradient norm was clipped at 1.0. The coefficients of the load-balancing and router $z$-loss terms were set to $\lambda_{\mathrm{bal}}=10^{-2}$ and $\lambda_z=10^{-3}$, respectively.

The comparative study included MeshGraphNets~\cite{pfaff2021meshgraphnets}, Transolver++~\cite{wu2025transolverpp}, DeepONet~\cite{lu2021deeponet}, and BlendedNet~\cite{sung2025blendednet}, representing graph-based message passing, attention-based neural operators, branch--trunk operator learning, and geometry-conditioned aerodynamic-field prediction, respectively. Each method retained its native representation mechanism, while the training--test partition, target normalization, supervised field data, and complete-surface test locations were held fixed. MeshGraphNets, Transolver++, DeepONet, BlendedNet, and GeoCANO contained $2.18$, $2.30$, $2.36$, $3.04$, and $3.47$ million trainable parameters, respectively. The comparison therefore controlled the training data and evaluation coordinates while preserving the architectural characteristics of the individual methods.

Field-prediction accuracy was evaluated separately for $C_p$, $\rho$, and $T$ using the mean absolute error (MAE), root mean square error (RMSE), and relative $L_2$ error. For each physical quantity, the metrics were first computed over the complete archived surface of each test configuration, assigning equal weight to every configuration. All field metrics were evaluated after inverse transformation to the original output scale, and the predicted values were not clipped. In addition to the pointwise field errors, the pressure-drag coefficient $C_{D,p}$ was compared with its CFD reference value.

\FloatBarrier

\subsection{Comparison with baseline methods}
\label{sec:overall-accuracy}

We compared the proposed method with the baseline models on all test configurations. Complete-surface metrics first quantified the accuracy of the individual predicted fields and the drag coefficient recovered from the predicted pressure field. Configuration-wise error distributions were then examined to determine whether the improvement persisted across the test set. Finally, spatial and gradient-conditioned diagnostics assessed the reconstruction of localized regions with rapid pressure variation.

Table~\ref{tab:baseline-physical-results} reports the complete-surface errors separately for $C_p$, $\rho$, and $T$, together with the errors in the pressure-drag coefficient $C_{D,p}$. For the three surface fields, each metric was evaluated over the complete archived surface of an individual configuration and then averaged across the test set. Here, $C_{D,p}$ is recovered by integrating the predicted $C_p$ field and therefore represents only the pressure-induced contribution to drag.

\begin{table}[!htbp]
    \centering
    \caption{Complete-surface field accuracy and drag-coefficient recovery over the test set.}
    \label{tab:baseline-physical-results}
    \scriptsize
    \setlength{\tabcolsep}{2.0pt}
    \resizebox{\textwidth}{!}{%
    \begin{tabular}{lcccccccccccc}
        \toprule
        & \multicolumn{3}{c}{$C_p$}
        & \multicolumn{3}{c}{$\rho$}
        & \multicolumn{3}{c}{$T$}
        & \multicolumn{3}{c}{$C_{D,p}$} \\
        \cmidrule(lr){2-4}
        \cmidrule(lr){5-7}
        \cmidrule(lr){8-10}
        \cmidrule(lr){11-13}
        Method
        & MAE & RMSE & Rel (\%)
        & MAE & RMSE & Rel (\%)
        & MAE & RMSE & Rel (\%)
        & MAE & RMSE & Rel (\%) \\
        \midrule
        DeepONet
        & $0.04623$ & $0.10511$ & $25.838$
        & $0.01762$ & $0.04010$ & $4.416$
        & $0.00127$ & $0.00310$ & $0.278$
        & $1.3299\times10^{-3}$ & $2.2674\times10^{-3}$ & $8.9055$ \\
        BlendedNet
        & $0.00384$ & $0.01115$ & $2.741$
        & $0.00146$ & $0.00427$ & $0.470$
        & $0.00024$ & $0.00055$ & $0.049$
        & $3.7078\times10^{-5}$ & $4.9127\times10^{-5}$ & $0.1929$ \\
        Transolver++
        & $0.00398$ & $0.01248$ & $3.067$
        & $0.00151$ & $0.00471$ & $0.519$
        & $0.00018$ & $0.00057$ & $0.051$
        & $4.9373\times10^{-5}$ & $7.0615\times10^{-5}$ & $0.2773$ \\
        MeshGraphNets
        & $0.00603$ & $0.02087$ & $5.130$
        & $0.00229$ & $0.00795$ & $0.875$
        & $0.00020$ & $0.00069$ & $0.062$
        & $9.3989\times10^{-5}$ & $1.1189\times10^{-4}$ & $0.4394$ \\
        \textbf{GeoCANO}
        & $\mathbf{0.00085}$ & $\mathbf{0.00675}$ & $\mathbf{1.661}$
        & $\mathbf{0.00033}$ & $\mathbf{0.00252}$ & $\mathbf{0.277}$
        & $\mathbf{0.00005}$ & $\mathbf{0.000255}$ & $\mathbf{0.023}$
        & $\mathbf{1.3552\times10^{-5}}$ & $\mathbf{2.2883\times10^{-5}}$ & $\mathbf{0.0898}$ \\
        \bottomrule
    \end{tabular}%
    }
\end{table}

The proposed method attained the lowest error for every physical quantity and evaluation metric in Table~\ref{tab:baseline-physical-results}. Relative to the best competing result for each metric, GeoCANO reduced the MAE by $77.9\%$, $77.4\%$, and $72.2\%$ for $C_p$, $\rho$, and $T$, respectively. The corresponding RMSE reductions ranged from $39.5\%$ to $53.6\%$, while the relative $L_2$ reductions ranged from $39.4\%$ to $53.1\%$. Among the three surface fields, $C_p$ retained the largest relative error, reflecting the greater difficulty of resolving localized steep pressure variations in the transonic regime. The improvement in the pressure field was also retained after surface integration. Relative to the best baseline result for each metric, GeoCANO reduced the MAE, RMSE, and relative $L_2$ error of $C_{D,p}$ by at least $53\%$. This result shows that the gain in field-level pressure accuracy translates into more accurate pressure-drag recovery.

\begin{figure}[!htbp]
    \centering
    \includegraphics[width=0.95\linewidth]{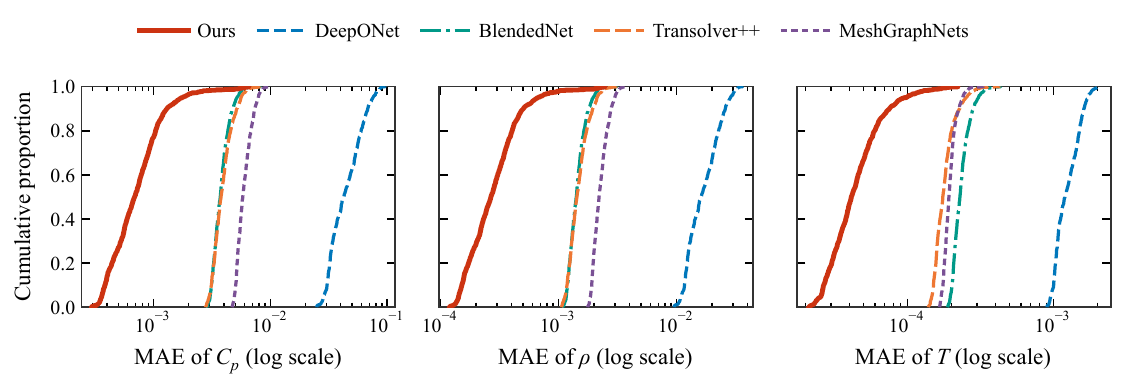}
    \caption{Empirical cumulative distributions of the configuration-wise MAE for $C_p$, $\rho$, and $T$.}
    \label{fig:baseline-error-distribution}
\end{figure}

Figure~\ref{fig:baseline-error-distribution} complements the mean field errors with the configuration-wise MAE distributions. For all three predicted fields, the distributions obtained by the proposed method were consistently shifted toward lower error relative to the baseline models, with the separation maintained over most of the test population.

\begin{figure}[!htbp]
    \centering
    \includegraphics[width=0.95\linewidth]{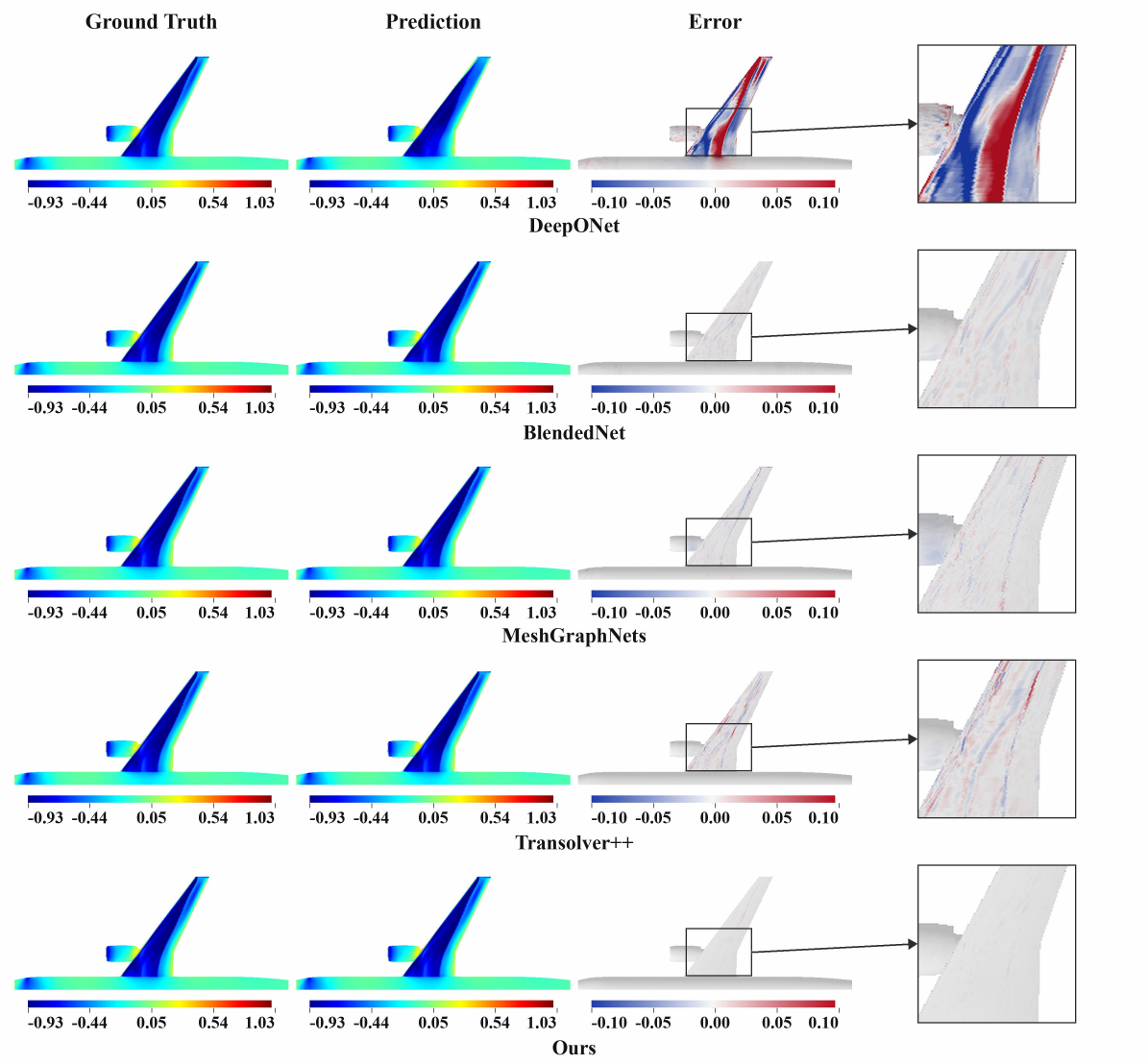}
    \caption{Surface-pressure reconstruction and residual morphology for a representative test case. Enlarged views highlight localized discrepancies in regions of rapid pressure variation.}
    \label{fig:local-pressure-reconstruction}
\end{figure}

For transonic surface fields, extensive smooth regions coexist with localized regions of rapid pressure variation, and configuration-level error statistics do not fully reveal the spatial organization of the error. Figure~\ref{fig:local-pressure-reconstruction} therefore examines the pressure reconstruction and residual morphology for a representative test configuration. The dominant discrepancies are concentrated primarily over the wing and in the wing--body interaction region, whereas substantially smaller errors occur over smoother portions of the surface. The enlarged views further show pronounced residual structures in regions of rapid pressure variation for the baseline models, whereas GeoCANO substantially attenuates these errors.

\begin{figure}[!htbp]
    \centering
    \includegraphics[width=0.65\linewidth]{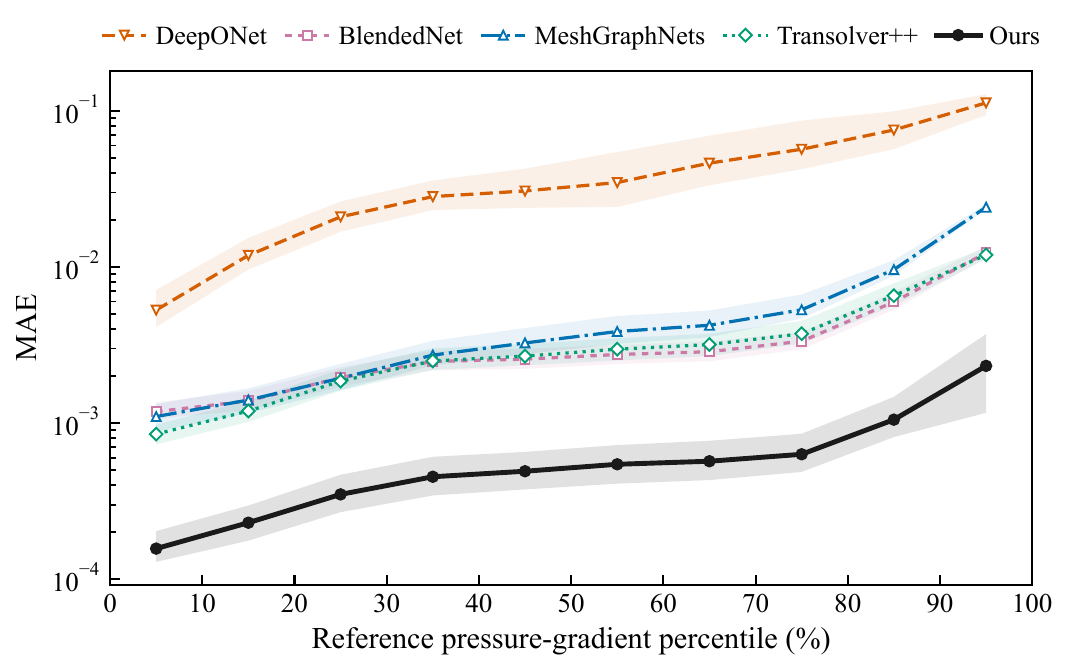}
    \caption{Gradient-conditioned $C_p$ prediction error over the test set. Within each configuration, surface nodes are ranked in descending order of the local CFD-reference $C_p$-gradient magnitude and partitioned into ten equal-population bins.}
    \label{fig:baseline-gradient-error}
\end{figure}

To evaluate this behavior systematically, Fig.~\ref{fig:baseline-gradient-error} reports the pressure error conditioned on the local variation of the CFD reference field. For each test configuration, the surface nodes were ranked in descending order according to the magnitude of the local reference $C_p$ gradient and divided into ten equal-population bins. The $C_p$ error was then computed independently within each bin before the configuration-level results were aggregated over the test set. This procedure separated the extensive smooth portions of the surface from the progressively more demanding regions of strong pressure variation without using the predicted field to define the groups. All methods exhibited larger errors in the higher-gradient bins, confirming that rapid pressure variations remain the most difficult part of the surface-field reconstruction problem. The proposed method nevertheless maintained the lowest error throughout the ten gradient-conditioned groups. Its advantage over the baseline models was therefore not confined to the smooth majority of the surface but persisted in the localized regions that dominate the reconstruction difficulty, consistent with the residual structures observed in Fig.~\ref{fig:local-pressure-reconstruction}.

\subsection{Coordinate encoding for multiscale surface-field reconstruction}
\label{sec:coordinate-encoding-analysis}

To assess the effectiveness of the proposed coordinate encoding for multiscale surface-field reconstruction, we conducted an ablation study against three alternative coordinate representations to distinguish the effects of axis-wise feature decomposition, cross-axis coupling, and parameter scale on reconstruction performance. The first, \emph{direct coordinates}, supplies the normalized Cartesian coordinates to the decoder without an explicit coordinate encoding. The second, \emph{axis-decomposed hash encoding}, performs independent multiresolution hash lookups along the three coordinate axes without explicitly coupling the retrieved axis features. The third is a traditional approach that applies a joint multiresolution hash lookup to the complete query coordinate. The proposed \emph{multiresolution hash-indexed canonical-polyadic (CP) coordinate representation} retrieves one-dimensional axis factors through independent hash mappings, combines them through explicit cross-axis products, and reweights the coupled rank components using the geometry latent. This design reduces the parameter cost of complex spatial encoding while maintaining reconstruction accuracy.

\begin{figure}[!htbp]
    \centering
    \includegraphics[width=0.88\linewidth]
    {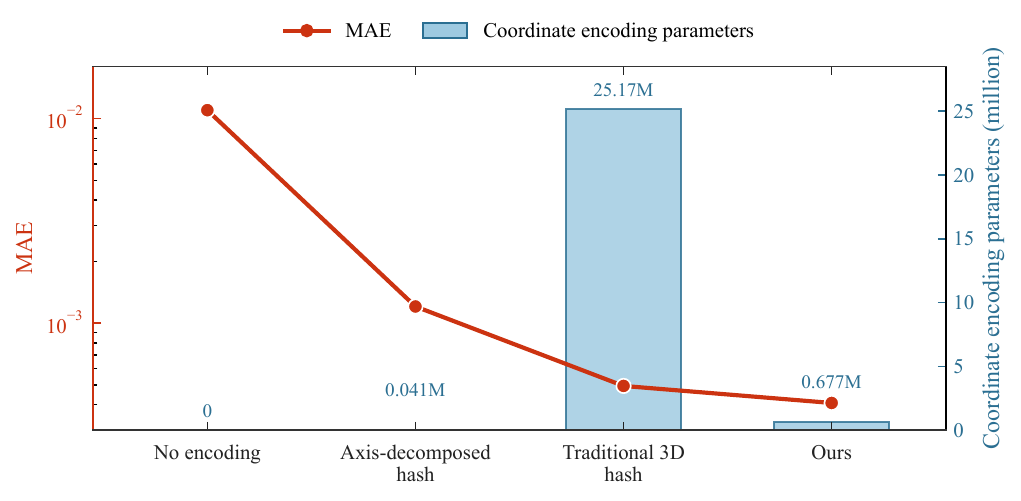}
    \caption{Aggregate complete-surface MAE and number of coordinate encode part trainable parameters for the four coordinate representations. Aggregate MAE is the arithmetic mean of the channel-wise $C_p$, $\rho$, and $T$ MAEs.}
    \label{fig:encoding-accuracy-parameters}
\end{figure}

Figure~\ref{fig:encoding-accuracy-parameters} and Table~\ref{tab:encoding-channelwise} compare the four coordinate representations at the aggregate and variable-wise levels, respectively. The aggregate MAE in the figure is the arithmetic mean of the $C_p$, $\rho$, and $T$ MAEs reported in the table. The parameter counts in Fig.~\ref{fig:encoding-accuracy-parameters} refer exclusively to the trainable parameters introduced by the coordinate-encoding module rather than the total model size. Direct coordinates introduced no trainable encoding parameters and yielded an aggregate MAE of $1.10\times10^{-2}$. Axis-decomposed hash encoding reduced the aggregate MAE to $1.21\times10^{-3}$ using $0.041$ million encoding parameters, whereas traditional three-dimensional hash encoding achieved $5.10\times10^{-4}$ but required $25.17$ million encoding parameters. The proposed CP representation attained the lowest aggregate MAE of $4.10\times10^{-4}$ with $0.677$ million encoding parameters. It therefore used only $2.7\%$ of the encoding parameters required by traditional three-dimensional hash encoding while further reducing the aggregate MAE by approximately $19.6\%$. Table~\ref{tab:encoding-channelwise} shows that the CP representation achieved MAEs of $8.5\times10^{-4}$, $3.3\times10^{-4}$, and $5.0\times10^{-5}$ for $C_p$, $\rho$, and $T$, respectively. The best result was obtained for all three output fields, indicating that the improvement was not dominated by a single physical quantity but remained consistent across the predicted surface fields.

\begin{table}[!htbp]
    \centering
    \caption{Variable-wise complete-surface MAE for the coordinate-representation ablation over the test configurations.}
    \label{tab:encoding-channelwise}
    \small
    \setlength{\tabcolsep}{8pt}
    \begin{tabular}{lccc}
        \toprule
        Coordinate representation
        & $C_p$
        & $\rho$
        & $T$ \\
        \midrule
        Direct coordinates
        & 0.022678
        & 0.009474
        & 0.000851 \\
        Axis-decomposed hash
        & 0.002544
        & 0.000961
        & 0.000119 \\
        Traditional 3D hash
        & 0.001057
        & 0.000417
        & 0.000057 \\
        Proposed CP representation
        & \textbf{0.000850}
        & \textbf{0.000330}
        & \textbf{0.000050} \\
        \bottomrule
    \end{tabular}
\end{table}

\begin{figure}[!htbp]
    \centering
    \includegraphics[width=0.95\linewidth]
    {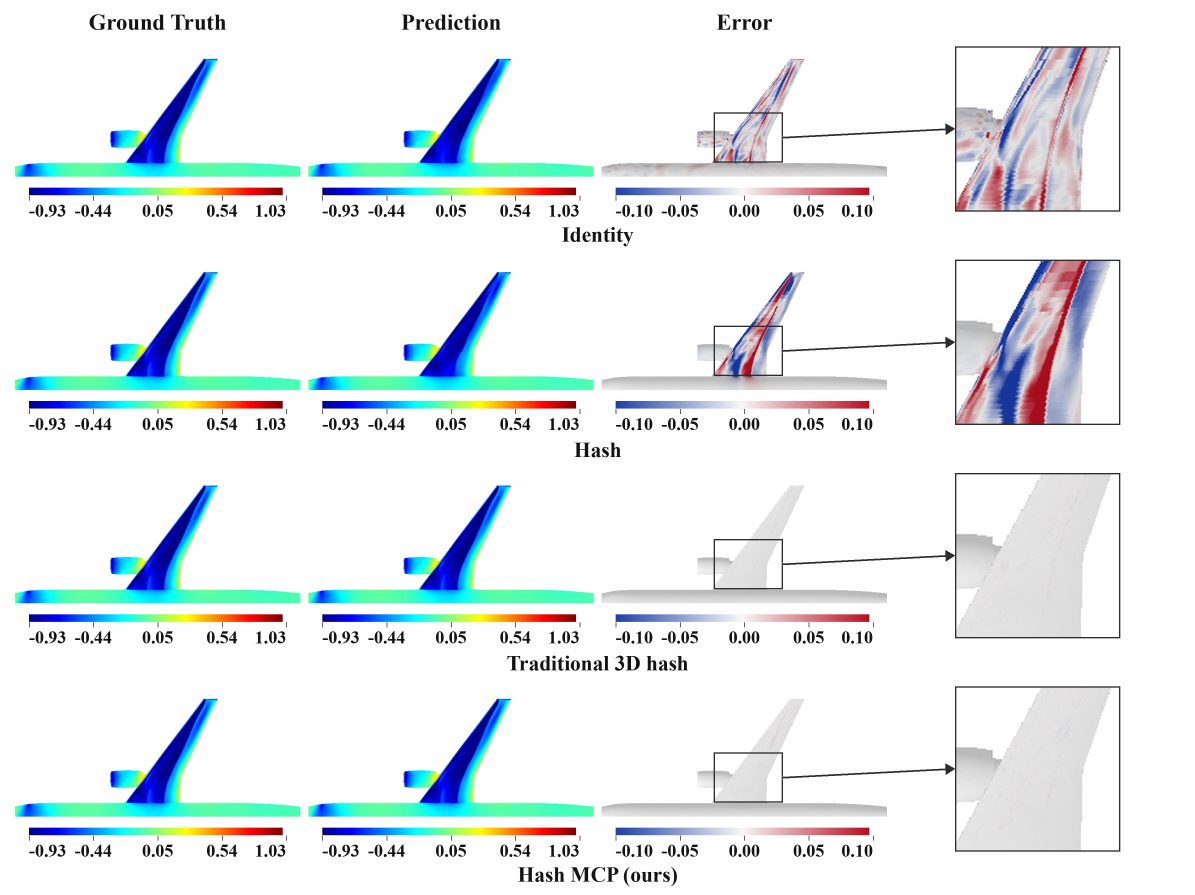}
    \caption{Surface-pressure reconstruction and residual morphology for a test configuration using direct coordinates, axis-decomposed hash encoding, Traditional 3D hash and the proposed multiresolution CP representation.}
    \label{fig:encoding-surface-comparison}
\end{figure}

The complete-surface metrics do not fully reveal how the encoding choice affects the spatial distribution of the error. Figure~\ref{fig:encoding-surface-comparison} therefore compares the four coordinate representations on a representative test configuration. Direct coordinates left extended high-amplitude residual bands over the wing and near the wing--body junction. Axis-decomposed hash encoding substantially suppressed the broad discrepancies, although localized residuals remained in regions of rapid pressure variation. The proposed CP representation further reduced both the magnitude and spatial extent of these residuals.

\begin{figure}[!htbp]
    \centering
    \includegraphics[width=0.65\linewidth]
    {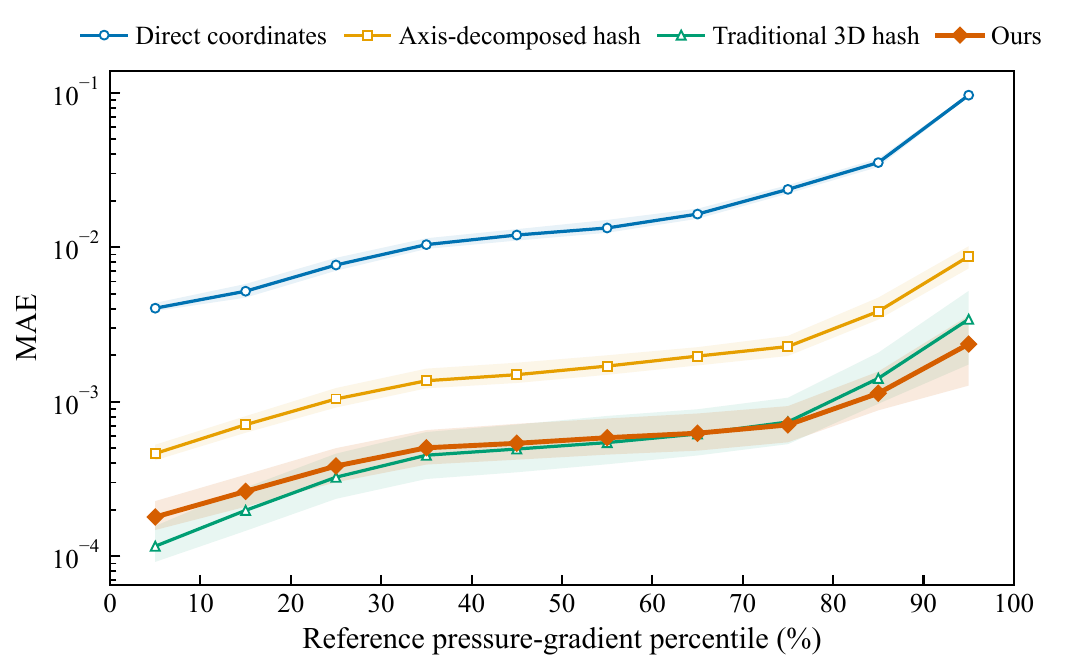}
    \caption{Gradient-conditioned $C_p$ prediction error for the four coordinate representations over the test configurations.}
    \label{fig:encoding-gradient-error}
\end{figure}

To assess the effect of parameter compression on prediction accuracy in locally complex flow regions, we applied the same reference-gradient grouping procedure used above. Figure~\ref{fig:encoding-gradient-error} reports the resulting gradient-conditioned errors. As the reference pressure gradient increased, all four coordinate representations exhibited increasing prediction error. Direct coordinate input deteriorated most strongly, whereas axis-decomposed hash encoding effectively reduced the error across the entire gradient range. Traditional three-dimensional hash encoding and the proposed CP representation maintained comparably low errors in the high-gradient groups, indicating that the CP representation preserved the local expressive capacity required for complex pressure variations. Its parameter compression was therefore not achieved at the expense of prediction accuracy in high-gradient regions.

The ablation results showed that multiresolution coordinate features substantially improved surface-field reconstruction relative to direct coordinate input, while explicit cross-axis coupling was important for further reducing the prediction error. The proposed CP representation achieved the best result for all three predicted quantities and obtained a lower aggregate MAE. In high-pressure-gradient regions, the CP representation effectively captured multiscale surface fields while substantially reducing model storage and training costs.

\FloatBarrier
\subsection{Routing structure in the sparse decoder}
\label{sec:expert-analysis}

With the coordinate representation fixed, we examined whether the query-wise sparse decoder developed structured expert contributions and systematically adjusted their allocation with local reconstruction difficulty. The router activates two of four experts at each query. Every expert predicts all three output fields, without predefined assignments to physical variables or aerodynamic regions during training. The analysis below focuses on the pressure channel. For any query, the complete pressure prediction is obtained as the routing-weighted sum of the pressure predictions from the two activated experts. Here, the active expert pair denotes the two experts selected at that query, and the routing weights are their normalized contribution coefficients.

Routing statistics over a fixed diagnostic subset showed a clearly nonuniform expert-utilization pattern without collapse to a single branch. E2 and E3 received $69.61\%$ of the total routing weight, and one of them was dominant at $80.04\%$ of the queries. By contrast, E4 had the highest selection rate, at $59.57\%$, but received a relatively small average weight when activated. E1 was selected at $35.79\%$ of the queries but was rarely dominant. Selection frequency and effective weighted influence were therefore not equivalent.

\begin{figure}[!htbp]
    \centering
    \includegraphics[width=0.95\linewidth]
    {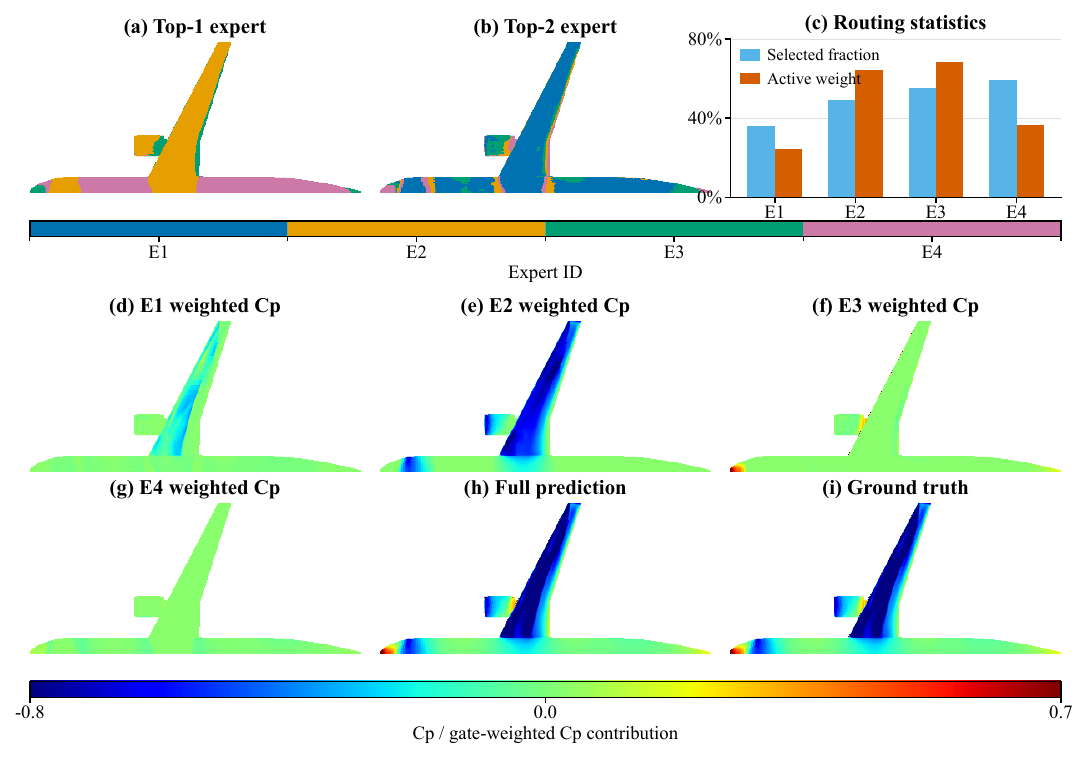}
    \caption{Spatial organization of the sparse decoder for an illustrative test configuration. Panels (a) and (b) identify the dominant and secondary experts, and panel (c) summarizes their selection fractions and conditional routing weights when activated. Panels (d)--(g) show the routing-weighted pressure-coefficient contributions, followed by the full-mixture prediction and CFD reference in panels (h) and (i).}
    \label{fig:expert-functional-differentiation}
\end{figure}

Figure~\ref{fig:expert-functional-differentiation} resolves this behavior spatially. The spatial distributions of the dominant and secondary experts partition the aircraft surface into coherent routing regions with partial overlap between different experts. The gate-weighted contribution fields show that E2 is primarily responsible for reconstructing the large-amplitude pressure distribution over the lifting surface, whereas the remaining experts provide complementary contributions over different spatial regions. These quantities represent the weighted contributions of individual experts within the trained mixture rather than their predictive capabilities when operating independently.

\begin{figure}[!htbp]
    \centering
    \includegraphics[width=0.95\linewidth]
    {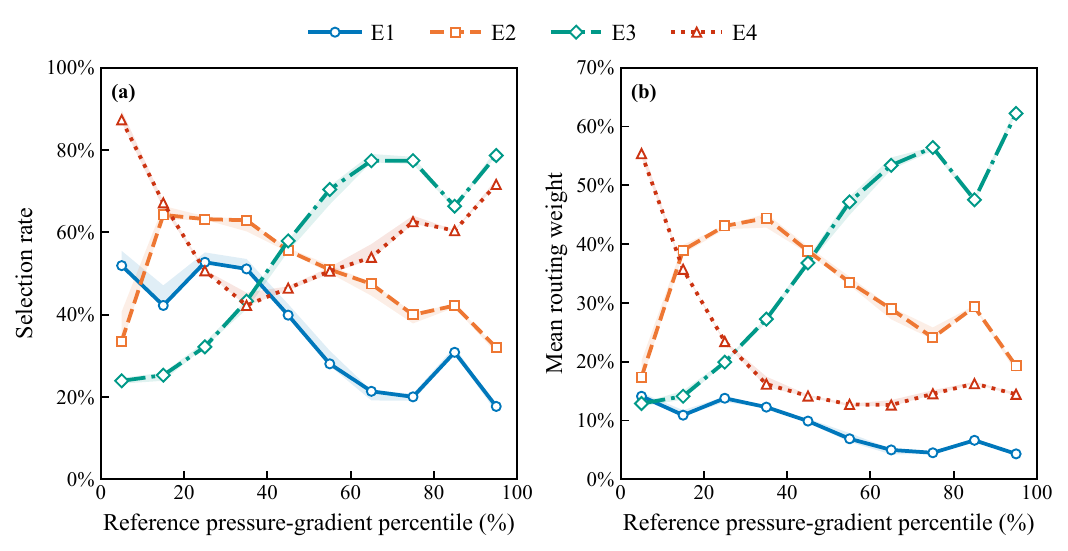}
    \caption{Gradient-conditioned expert allocation over the test configurations. Panel (a) reports the fraction of queries for which each expert is selected, and panel (b) reports its mean routing weight over all queries in the corresponding gradient interval.}
    \label{fig:expert-gradient-routing}
\end{figure}

To determine whether this allocation varied systematically with local field complexity, we used the CFD-reference pressure-gradient indicator introduced in Section~\ref{sec:overall-accuracy}. Figure~\ref{fig:expert-gradient-routing} reports the selection rates and mean routing weights within each gradient interval. The selection rate records how often an expert enters the active pair, whereas the mean routing weight is computed over all queries in the interval, including the zero weights of inactive experts. The selection rate therefore characterizes the spatial participation of an expert, while the mean routing weight measures its effective mixture contribution over the complete gradient interval. The gradient-conditioned routing results revealed a progressive redistribution of the mixture with local pressure variation. In the lowest-gradient interval, E4 was selected at approximately $88\%$ of the queries and received a mean routing weight of about $55\%$. E2 was most prominent over the low-to-intermediate gradient range, where its selection rate and mean routing weight reached approximately $65\%$ and $44\%$, respectively. With increasing pressure gradient, both quantities rose markedly for E3. In the highest-gradient interval, E3 was selected at approximately $79\%$ of the queries and carried a mean routing weight of about $62\%$. By contrast, E1 became less involved as the gradient increased. Although E4 was again frequently selected in the highest-gradient interval, its mean routing weight remained near $15\%$, indicating that it mainly acted as the secondary member of the active pair.

These results show that the routing allocation is not only spatially organized but also conditioned on local reconstruction difficulty. Routing mass shifts toward E3 in regions of stronger pressure variation, whereas E2 and E4 make more prominent contributions in the low- and intermediate-gradient regions.

\begin{figure}[!htbp]
    \centering
    \includegraphics[width=0.95\linewidth]
    {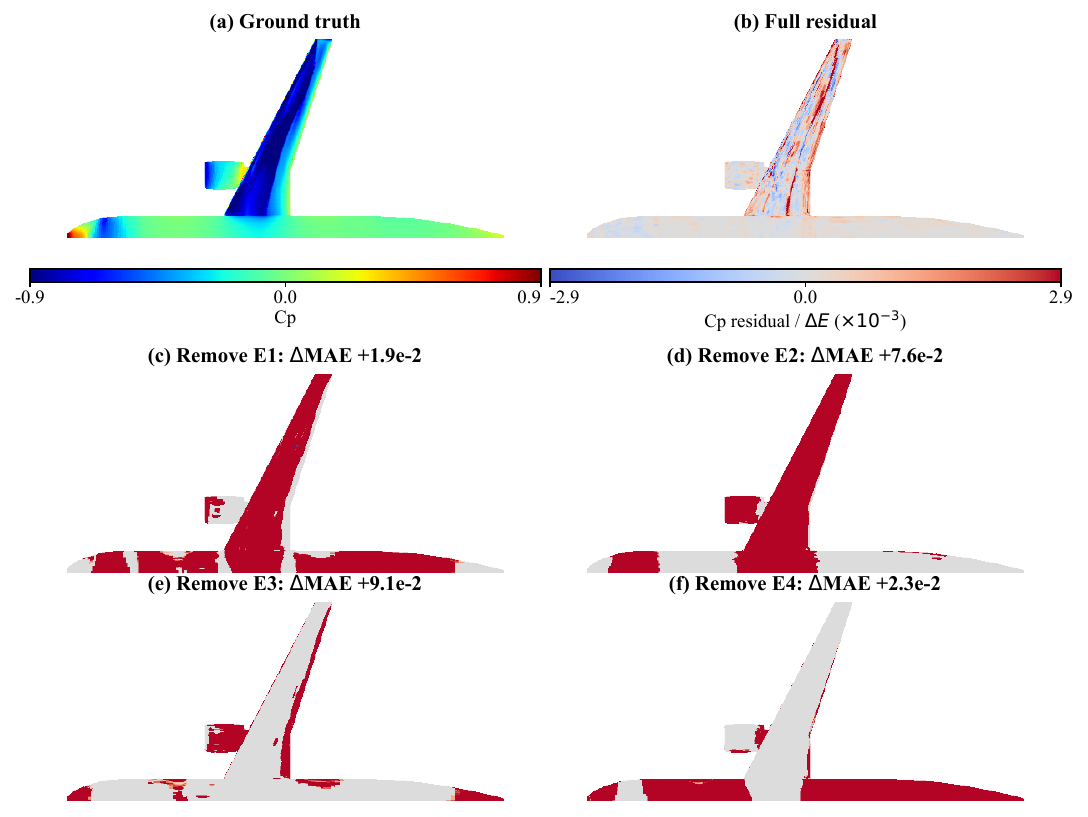}
    \caption{Leave-one-expert-out intervention in the trained sparse decoder. The CFD reference and intact-mixture residual are followed by the residual fields obtained after removing each routed expert contribution.}
    \label{fig:expert-removal}
\end{figure}

The routing diagnostics describe the allocation pattern but do not directly establish whether the resulting expert contributions materially affect the reconstructed field. We therefore removed the weighted contribution of one expert from the mixture prediction at a time while preserving the original contributions of the remaining experts and the original routing decisions. No rerouting or weight renormalization was applied. Figure~\ref{fig:expert-removal} reports the resulting residual fields. The degradation measures the operational dependence of the trained mixture on the corresponding expert contribution, rather than the standalone capability or strict nonreplaceability of that expert.

Removing E3 produced the largest increase in pressure-coefficient MAE, approximately $9.1\times10^{-2}$, with pronounced degradation in localized pressure-sensitive regions. Removing E2 yielded an increase of approximately $7.6\times10^{-2}$ and a broader deterioration over the lifting surface, consistent with its large-amplitude contribution in Fig.~\ref{fig:expert-functional-differentiation}. The corresponding increases for E4 and E1 were smaller, approximately $2.3\times10^{-2}$ and $1.9\times10^{-2}$, but their residual patterns remained spatially distinct. The disproportionate effect of removing E3 is consistent with its increased routing weight in the highest-gradient interval, whereas the broad effect of removing E2 reflects its important role across the low- and intermediate-gradient regions.

In summary, expert routing in the sparse decoder exhibits three characteristics: different experts occupy structured spatial routing regions, their allocation changes with the local pressure gradient, and their weighted contributions affect the final prediction to different degrees. These results show that query-wise sparse routing can develop an implicit functional division adapted to local flow-field complexity without predefined physical assignments for the experts.

\FloatBarrier
\subsection{Off-topology coordinate queries within and beyond the training distribution}
\label{sec:off-topology-queries}
\label{sec:off_topology_near_wall}

The continuous-coordinate interface was evaluated at locations that were not
restricted to the discrete topology presented to each model. Two regimes were
distinguished. The first queried omitted coordinates on the aircraft surface
and therefore remained within the surface domain represented during training.
The second displaced the queries from the surface into previously unseen
off-wall layers and constituted an out-of-distribution evaluation. This
separation distinguishes robustness to changes in output topology from
extrapolation beyond the training surface.

\paragraph{\textbf{In-distribution queries on the aircraft surface.}}
The surface of each test configuration was progressively simplified into four
strictly nested query topologies with retention ratios of $1/2$, $1/4$, $1/8$,
and $1/16$. At each level, the retained nodes defined the explicit topology,
while the omitted nodes served as off-topology query locations on the same
aircraft surface. The nested topologies were constructed through balanced
contractions of adjacent edges using mesh connectivity and training-set
geometry. GeoCANO, DeepONet, BlendedNet, and Transolver++ directly evaluated
both retained and omitted coordinates. MeshGraphNets instead performed
inference on the simplified graph and reconstructed the omitted values through
harmonic interpolation on the original surface mesh.

\begin{figure}[!htbp]
    \centering
    \includegraphics[width=0.98\textwidth]
    {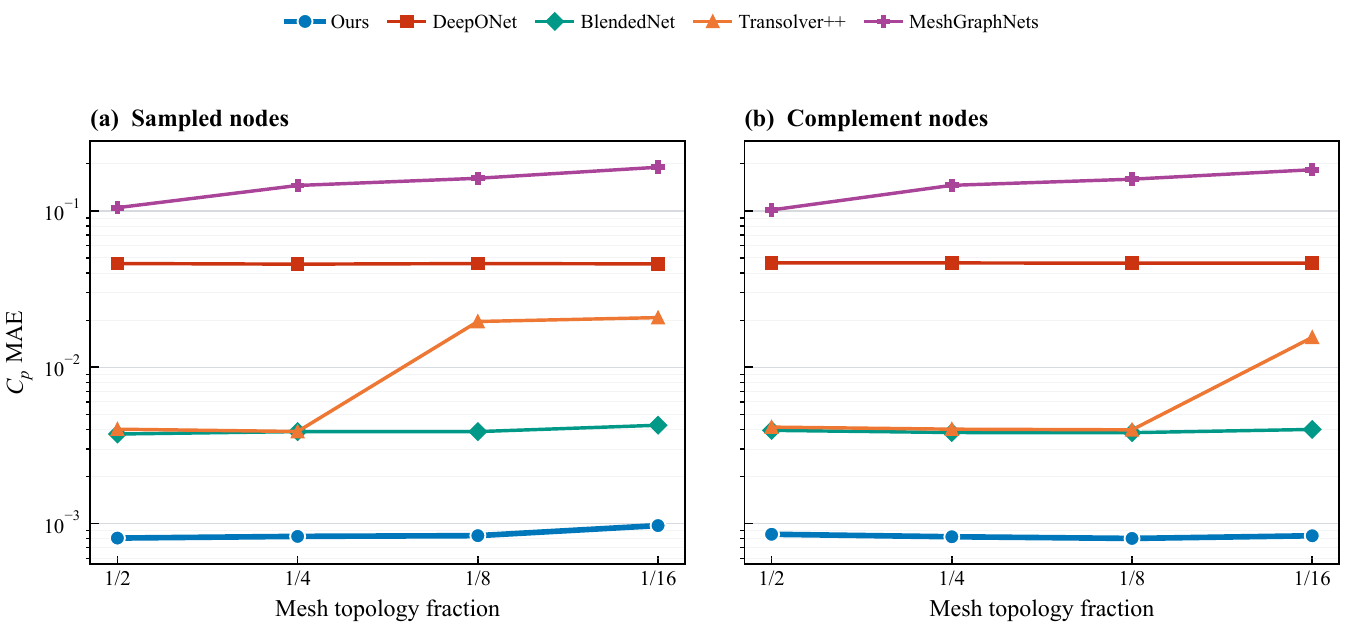}
    \caption{In-distribution $C_p$ MAE at retained and off-topology surface
    coordinates under different query topologies.}
    \label{fig:topology-sampled-complement}
\end{figure}

Figure~\ref{fig:topology-sampled-complement} shows that GeoCANO maintained
consistent full-surface accuracy as the explicit topology became sparser,
and achieved the lowest MAE at every retention ratio. DeepONet and BlendedNet
were also relatively insensitive to the topology change, but their errors
remained substantially higher. Transolver++ deteriorated at the two sparsest
levels, while the error of MeshGraphNets increased continuously as graph
simplification and interpolation became more pronounced. GeoCANO also
produced comparable errors at retained and omitted surface coordinates,
indicating that its in-distribution accuracy is not tied to a particular node
organization.

\begin{figure}[!htbp]
    \centering
    \includegraphics[width=0.98\textwidth]
    {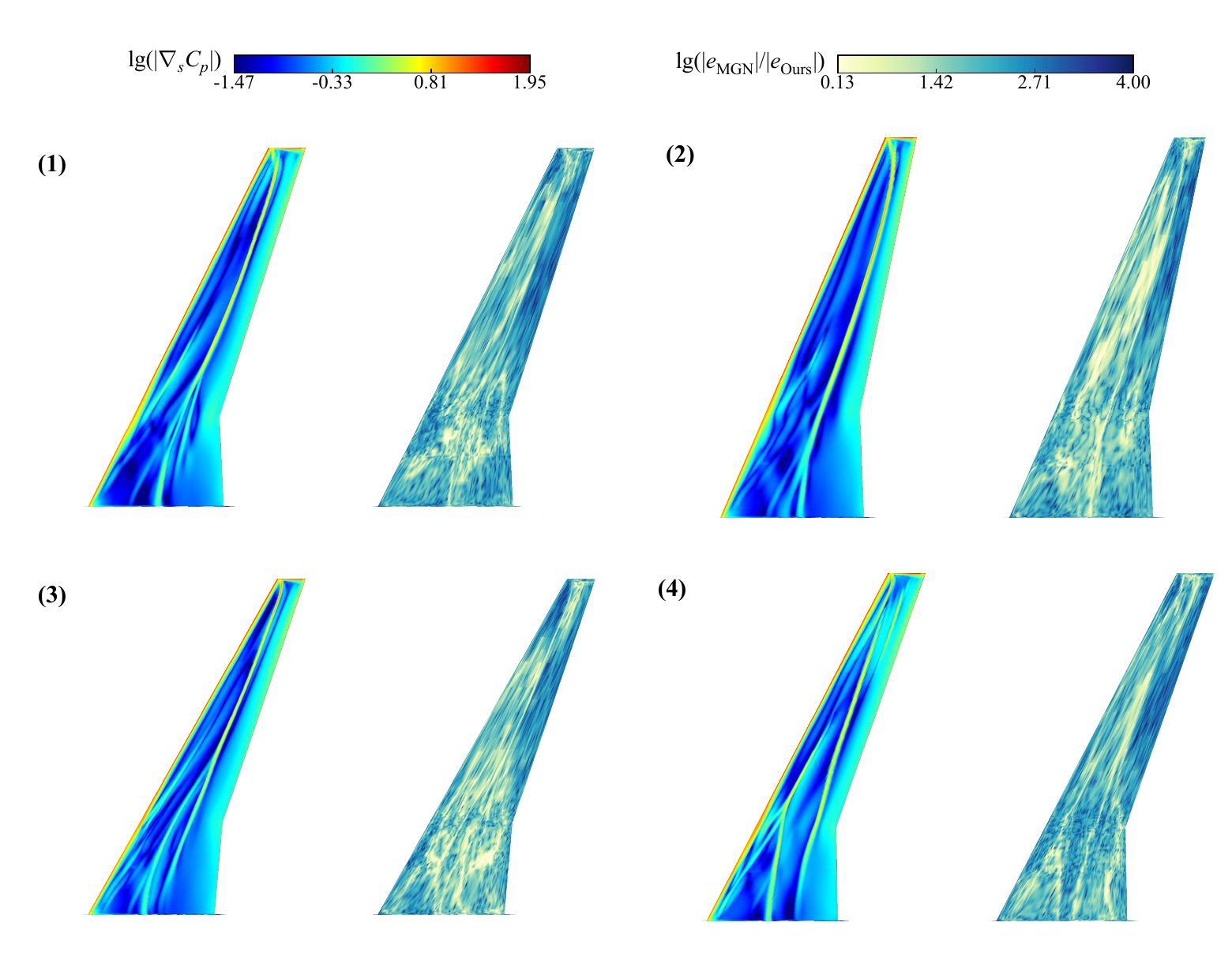}
    \caption{Surface-pressure-gradient magnitude and local error ratio for
    representative in-distribution off-topology test cases under the sparsest
    query topology. Each paired group corresponds to one test configuration:
    the left panel shows the CFD-reference gradient magnitude, and the right
    panel shows the MeshGraphNets-to-GeoCANO ratio of pointwise absolute
    $C_p$ errors. Ratios above unity favor GeoCANO.}
    \label{fig:topology-gradient-advantage}
\end{figure}

Figure~\ref{fig:topology-gradient-advantage} localizes this difference under
the sparsest query topology. Each paired group corresponds to one
representative test configuration: the left panel shows the magnitude of the
CFD-reference surface-pressure gradient, whereas the right panel shows the
pointwise absolute-error ratio between MeshGraphNets and GeoCANO. The ratio is
defined as the MeshGraphNets error divided by the GeoCANO error, so values
above unity indicate a local advantage for GeoCANO. This advantage extended
over the main wing and became more pronounced near the leading edge, rapidly
varying pressure regions, and the wing--body junction. Its spatial
distribution partially followed the surface-pressure gradient, indicating that
rapid field variation amplified the smoothing and propagation errors
introduced by graph coarsening and subsequent interpolation. Direct
coordinate evaluation better preserved the localized pressure structures at
omitted surface coordinates.

\FloatBarrier

\paragraph{\textbf{Out-of-distribution queries at off-wall locations.}}
The second regime moved the queries away from the aircraft surface to examine the extent to which surface-trained models could be evaluated beyond their supervised coordinates. All models were evaluated without fine-tuning at successive CFD layers. The surface input was fixed at a sampling ratio of $1/16$, and the connectivity required by topology-dependent methods was reconstructed from the retained surface points. Let $d$ denote the query-to-wall distance and $d_1$ the distance of the first off-wall layer. The evaluated locations spanned $d/d_1=2^1$ to $2^8$, with all methods queried at identical coordinates and compared against the same CFD references.

\begin{figure}[htbp]
\centering
\includegraphics[width=0.88\linewidth]
{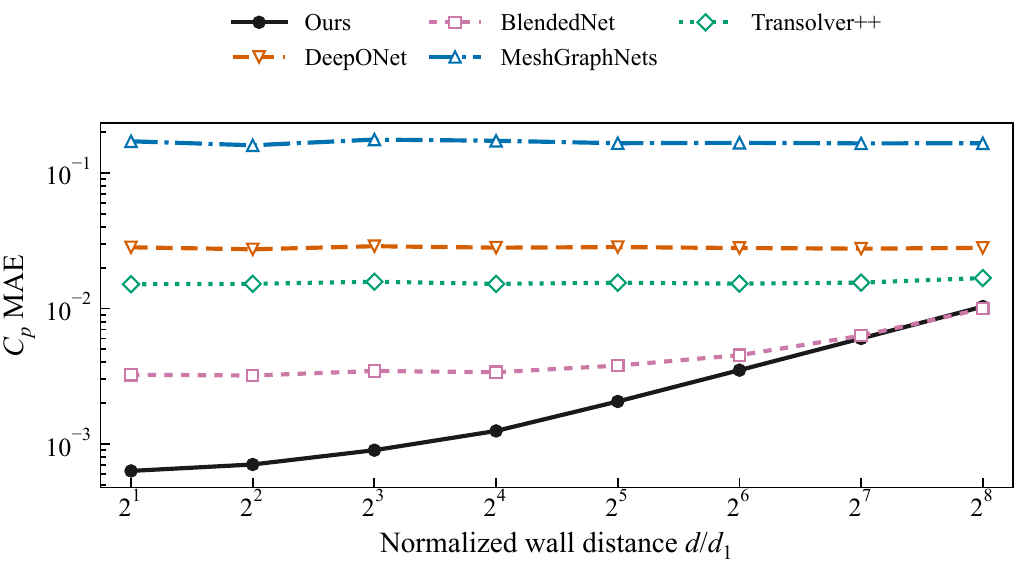}
\caption{Out-of-distribution $C_p$ MAE at successive off-wall layers using a surface sampling ratio of $1/16$. The vertical axis is logarithmic.}
\label{fig:off_topology_distance}
\end{figure}

As shown in Fig.~\ref{fig:off_topology_distance}, GeoCANO consistently yielded lower errors than the compared methods. Its advantage was most pronounced in the innermost layers and gradually diminished as the queries moved farther from the surface. This distance-dependent behavior indicates that, although GeoCANO can be evaluated directly at off-wall coordinates, its accurate extrapolation was demonstrated only over a finite near-wall range. Within this range, the off-wall pressure field remained strongly correlated with the surface distribution, allowing the geometry-conditioned coordinate representation to extend the learned surface field over limited normal distances without volumetric supervision.

\FloatBarrier

\section{Conclusions}
\label{sec:conclusion}

This study formulated physical-field prediction over varying three-dimensional geometries as a function-learning problem conditioned on the complete geometry and introduced the Geometry-Conditioned Adaptive Neural Operator (GeoCANO). A global geometry state modulates a multiresolution axis-factorized coordinate representation to construct a configuration-specific field approximation space, while a query-wise sparse mixture-of-experts decoder allocates model capacity according to local approximation difficulty. By separating geometry encoding from coordinate-wise evaluation, GeoCANO encodes each configuration once and subsequently evaluates the corresponding field directly on different surface point sets and local regions without rebuilding the computational graph for each output topology.

Transonic aerodynamic-field prediction over complete aircraft configurations was considered as a representative test case. For unseen wing--body--nacelle configurations within the sampled design family, GeoCANO achieved lower errors than the considered baselines in both complete-surface prediction and regions characterized by rapid pressure variation. The improvement in pressure-field accuracy was retained after surface integration, resulting in more accurate recovery of the pressure-drag coefficient. The ablation studies showed that multiresolution features substantially improved field reconstruction relative to direct coordinate input. Compared with conventional three-dimensional hash encoding, the geometry-modulated axis-factorized representation provided a more favorable balance between prediction accuracy and parameter count. The routing diagnostics further showed that expert allocation varied with local pressure variation and that the predictions exhibited unequal sensitivity to individual expert contributions, indicating spatially structured functional specialization. Direct evaluation on successive off-wall layers also demonstrated that the learned representation remained effective over a finite near-wall range without retraining or intermediate interpolation, although its relative advantage decreased with increasing wall distance.

The present validation is limited to the wing--body--nacelle geometry family considered in this study. The applicability of GeoCANO to independently remeshed surfaces, topologically distinct geometries, and the complete surrounding flow domain remains to be established. Its generalization to other physical systems with varying domains also requires systematic assessment. Future work will extend the formulation to joint conditioning on geometry and operating parameters and investigate coupled surface--volume prediction on heterogeneous discretizations. Gradient-aware sampling and physically informed constraints will also be incorporated to improve the resolution of complex local structures and transferability across problems.

This work was supported by ZJU Kunpeng $\&$ Ascend Center of Excellence.

\bibliographystyle{elsarticle-num}
\bibliography{reference}

\end{document}